\documentstyle[preprint,aps,prl,amsmath]{revtex}

\begin{document}

\begin{center}
{\Large
\bf Direct Numerical Simulations of the Navier-Stokes Alpha Model}

\vspace{0.3in}

Shiyi Chen$^{1,2}$, Darryl D. Holm$^{1}$, Len G. Margolin$^{3}$
and Raoyang Zhang$^{1,4}$
\end{center}

\vspace{0.3in}

${}^{1}${\scriptsize Center for Nonlinear Studies and Theoretical Division,
Los Alamos National Laboratory, Los Alamos, NM 87545}

${}^{2}${\scriptsize IBM Research Division,
T.J. Watson Research Center, P.O. Box 218, Yorktown Heights, NY 10598}

${}^{3}${\scriptsize Applied Theoretical and Computational Physics Division,
Los Alamos National Laboratory, Los Alamos, NM 87545}

${}^{4}${\scriptsize Department of Mechanical Engineering,126 Spencer
Laboratory, University of Delaware, Newark, DE 19716 }

\date{}
\vspace{0.5in}
\begin{abstract}

We explore the utility of the recently proposed alpha equations in 
providing a subgrid model for fluid turbulence.  Our principal results are 
comparisons of direct numerical simulations of fluid turbulence using 
several values of the parameter alpha, including the limiting case where the 
Navier-Stokes equations are recovered.  Our studies show that the large 
scale features, including statistics and structures, are preserved by the alpha 
models, even at coarser resolutions where the fine scales are not fully 
resolved.  We also describe the differences that appear in simulations.  
We provide a summary of the principal features of the alpha equations, 
and offer some explanation of the effectiveness of these equations used 
as a subgrid model for three-dimensional fluid turbulence.

\end{abstract}

\vspace{0.3in}


\pagebreak

\section{Introduction}

Holm et al.~\cite{HMR[1998a]},~\cite{HMR[1998b]} introduced the ``alpha
models'' for the mean motion of ideal incompressible fluids as the
$n$-dimensional generalization of the one-dimensional Camassa-Holm
(CH) equation. The 1D CH equation describes shallow water waves with
nonlinear dispersion and admits soliton solutions called
``peakons''~\cite{CH[1993]}. Its $n$-dimensional generalization describes the
slow time dynamics of fluids in which nonlinear dispersion accounts for
the effects of the small scale rapid variability upon the mean motion. 
The fluid transport velocity is found by inversion of a Helmholtz
operator acting on fluid circulation (or momentum) velocity.
This operator contains a length scale that corresponds to the magnitude
of the fluctuation covariance; the application of this operator smoothes
the transport velocity relative to the circulation velocity.
This length scale is denoted by $\alpha$ in references~\cite{HMR[1998a]},
\cite{HMR[1998b]}, \cite{CH[1993]}, hence the name {\it alpha models} for
these mean fluid motion theories. The alpha models for self consistent
mean fluid dynamics are derived by applying temporal averaging
procedures to Hamilton's principle for an ideal incompressible fluid flow.
The resulting mean fluid motion equations are obtained by using the
Euler-Poincar\'e variational framework~\cite{HMR[1998a]},~\cite{HMR[1998b]}.
(Euler-Poincar\'e equations are the Lagrangian version of Lie-Poisson
Hamiltonian systems.) Therefore, these equations possess conservation laws for
energy and momentum, as well as a Kelvin-Noether circulation theorem that
establishes how the time average (or perhaps statistical) properties of the
fluctuations affect the circulation of the mean flow. These ideal
fluid equations also describe geodesic motion on the volume-preserving
diffeomorphism group for a metric containing the $H^1$ norm of the mean fluid
velocity. Their geometrical properties are discussed in~\cite{HKMRS[1998]}.
Their relation to Eulerian and Lagrangian mean fluid theories is discussed
in~\cite{Holm[1999]}. In recognition of their origins, these mean fluid
motion equations may be known equally well by either the name alpha models, or
CH equations.

Chen {\it et al.}~\cite{Chen etal[1998a]} -- \cite{Chen etal[1998c]}
introduced phenomenological viscosity into the CH equation and proposed the
resulting viscous Camassa-Holm equation (VCHE) as a closure approximation for
the Reynolds averaged equations of the incompressible Navier-Stokes fluid.
They tested this approximation on turbulent channel and pipe flows with steady
mean, by finding analytical solutions of the VCHE for the mean velocity and
the Reynolds shear stress and comparing them with
experiments~\cite{Zagarola96}. They found that the steady VCHE profiles
are consistent with data obtained from mean flow turbulence measurements in
most of the flow region for channels and pipes at moderate to high Reynolds
numbers. Thus, Chen {\it et al.} demonstrated a connection between
turbulence and the VCHE for steady, or mean solutions. In fact, the {\it
time-dependent} VCHE in a periodic box has unique classical solutions and a
global attractor whose fractal dimension is finite and scales according to
Kolmogorov's estimate, $N\sim(L/\ell_d)^3$, where
$\ell_d=(\nu^3/\epsilon)^{1/4}$ is the Kolmogorov dissipation
length~\cite{FHT98}. We note that the time-dependent VCHE model is {\it not}
equivalent to the Navier-Stokes equations with hyperviscosity, see~\cite{Chen
etal[1998a]} -- \cite{Chen etal[1998c]},~\cite{FHT98}. The VCHE model 
is also known as the Navier-Stokes alpha model, or the viscous alpha model.

The purpose of this paper is to compare the statistics and
structures of the velocity and vorticity fields at moderate Reynolds numbers
in a direct numerical simulation (DNS) of the viscous alpha model
with the corresponding results for the Navier-Stokes equations.
Through this comparison, we hope to determine whether the Navier-Stokes
 alpha model can be used as a subgrid model for fluid turbulence. 
The paper is arranged as follows: the mathematical background of the viscous alpha
model and its connections  with the Navier-Stokes equations are given in the
next section. The remainder of the paper compares the classical Navier-Stokes
DNS turbulence results with those for the viscous alpha models. In Section III we
present the numerical methodology and energy spectra. Stretching dynamics,
including the vorticity  structure and the alignment phenomena are presented
in  Section IV. In Section V we present various two-point statistics, including the
probability density function (PDF) of the velocity increment and the velocity
structure functions. Concluding remarks are given in Section VI.


\section{Eulerian Mean Theory and Alpha Model Equations}

Holm, Marsden, and Ratiu~\cite{HMR[1998a]},~\cite{HMR[1998b]} used
variational asymptotics to obtain evolution equations for the Eulerian mean
hydrodynamic motion of ideal incompressible fluids, employing an approximation of
Hamilton's principle for Euler's equation in a Euclidean space setting.
The method assumes that the Euler flow may be decomposed into its mean and
fluctuating components at a fixed position in space. In their approach, a 
first-order Taylor expansion in the fluctuation amplitude is used to approximate the
velocity field with the result that the $L^2$ metric in the Hamilton's
principle giving rise to the Euler equations is replaced by an $H^1$ metric
that produces the evolution equation of the Eulerian mean flow.  We shall call
this evolution equation the Euler alpha model, or the $n$-dimensional 
CH equation.

Holm et al.~\cite{HKMRS[1998]} give a geometrically intrinsic (i.e., coordinate 
free) derivation of these averaged equations by the procedure of {\it variational
asymptotics\/}, namely, by deriving an averaged Lagrangian and using this
Lagrangian to generate the equations via Hamilton's principle. This intrinsic
setting is useful because many interesting flows, e.g.,  flows on spheres such as
those in geophysics, do occur on manifolds. The standard decomposition into
mean
and fluctuating components is an additive decomposition, only valid in the
presence of a vector space structure. We follow Holm~\cite{Holm[1999]} in
presenting the derivation in Euclidean space.

We develop the Euler-Poincar\'e theory of advected fluctuations from the
viewpoint of Eulerian averaging. Our point of departure is a Lagrangian
comprised of the fluid kinetic energy in the Eulerian description, in which
volume preservation is imposed by a Lagrange multiplier $P$ (the pressure),
\begin{equation}\label{Lag-Eul-A}
L(\omega)
=
\int d^{\,3}x \left\{\frac{D}{2}
|{\bf U}({\bf x},t\,;\omega)|^2
+
P({\bf x},t\,;\omega) \Big(1-
D({\bf x},t\,;\omega) \Big)\right\}\,,
\end{equation}
where $D$ is the Eulerian volume element. We assume that there are two time
scales for the motion: the fast time $\omega$ and the slow time $t$. The
traditional Reynolds decomposition of fluid velocity into its fast and
slow components is expressed at a given position ${\bf x}$ in terms of the
Eulerian mean fluid velocity
${\bf u}$ as
\begin{equation}
{\bf U}({\bf x},t\,;\omega)
\equiv
{\bf u}({\bf x},t)
+ {\bf u}^{\,\prime}({\bf x},t\,;\omega)
\,.
\label{Re-vel-decomp}
\end{equation}
Following Holm~\cite{Holm[1999]}, we assume the Eulerian velocity fluctuation
${\bf u}^{\,\prime}({\bf x},t\,;\omega)$ is related to an Eulerian
fluid parcel displacement fluctuation --- denoted as $\boldsymbol\zeta
({\bf x},t\,;\omega)$ --- by
\begin{equation}
\frac{\partial\boldsymbol\zeta}{\partial t}
+
{\bf u}\,\boldsymbol{\cdot\nabla\zeta}
=
\boldsymbol{\zeta\cdot\nabla} {\bf u}
+
{\bf u}^{\,\prime}({\bf x},t\,;\omega)
\,.\label{order-zeta-A}
\end{equation}
For purely Eulerian
velocity fluctuations as in Eq. (\ref{Re-vel-decomp}), this relation
separates into two relations: the ``Taylor-like'' hypothesis
of~\cite{HKMRS[1998]},
\begin{equation}
\frac{\partial\boldsymbol\zeta}{\partial t}
+
{\bf u}
\boldsymbol{\cdot\nabla\zeta}
=
0
\,;\label{Taylor-like-hypoth}
\end{equation}
and the relation~\cite{HKMRS[1998]},~\cite{Holm[1999]}
\begin{equation}
0
=
\boldsymbol{\zeta\cdot\nabla} {\bf u}
+
{\bf u}^{\,\prime}({\bf x},t\,;\omega)
\,.\label{u-prime-zeta}
\end{equation}
Hence, the Reynolds velocity decomposition (\ref{Re-vel-decomp})
separates the Lagrangian (\ref{Lag-Eul-A}) into its mean and fluctuating
pieces as
\begin{equation}\label{Lag-Eul-B}
L(\omega)
=
\int d^{\,3}x \left\{\frac{D}{2}
\big|{\bf u}({\bf x},t)
+
{\bf u}^{\,\prime}({\bf x},t\,;\omega)
\big|^2
+
P({\bf x},t) \Big(1-
D({\bf x},t) \Big)\right\}
\,.
\end{equation}
No modification is needed in the pressure constraint in this Lagrangian,
because
the Eulerian mean {\it preserves} the condition that the velocity be
divergence free; hence,
$\boldsymbol{\nabla\cdot}
{\bf u} = 0$. It remains only to take the Eulerian mean
of this Lagrangian, in which we assume
$\langle\boldsymbol\zeta\,\rangle = 0$.
The Eulerian mean averaging process at fixed position ${\bf x}$ is
denoted $\langle\,\boldsymbol{\cdot}\,\rangle$ with, e.g.,
\begin{equation}\label{Eul-mean-def}
{\bf u}({\bf x},t)
=
\langle{\bf U}({\bf x},t\,;\omega)\rangle
\equiv
\lim_{T\to\infty}\frac{1}{T}\int_{0}^T
{\bf U}({\bf x},t\,;\omega)\,d\omega\,.
\end{equation}
By Eq. (\ref{u-prime-zeta}), the Eulerian mean kinetic energy due to the
velocity fluctuation satisfies
\begin{equation}\label{Eul-mean-fluct-KE}
\langle\,|{\bf u}^{\,\prime}|^2\,\rangle
= \langle\zeta^k\zeta^l\rangle
{\bf u}_{,k}\boldsymbol\cdot{\bf u}_{,l}
\,.
\end{equation}
Thus, we find the following Eulerian mean Lagrangian,
\begin{equation}\label{Lag-Eul-B-1}
\langle{L}\rangle
=
\int d^{\,3}x \left\{\frac{D}{2}\Big[
\big|{\bf u}({\bf x},t)\big|^2
+
\langle\,\zeta^k\zeta^l\rangle
{\bf u}_{,k}
\boldsymbol{\cdot}
{\bf u}_{,l}
\Big]
+
P({\bf x},t) \Big(1-
D({\bf x},t) \Big)\right\}\;.
\end{equation}
The advection relation  (\ref{Taylor-like-hypoth}) implies the same
advective velocity for each component of the symmetric Eulerian mean covariance
tensor $\langle\,\zeta^k\zeta^l\rangle$. Thus, we have
\begin{equation}
\Big(
\frac{\partial}{\partial t}
+
{\bf u}
\boldsymbol{\cdot\nabla}
\Big)
\langle\,\zeta^k\zeta^l\rangle
=
0
\,.\label{cov-dyn-E}
\end{equation}
Together, this relation and the continuity equation for the volume element
$D$,
\begin{equation}
\frac{\partial{D}}{\partial t}
+
\boldsymbol{\nabla\,\cdot\,}
D{\bf u}
=
0
\,,\label{cont-eqn-A}
\end{equation}
complete the auxiliary equations needed for deriving the equation of motion for
the Eulerian mean velocity ${\bf u}$ from the averaged
Lagrangian $\langle{L}\rangle$ in (\ref{Lag-Eul-B}) by using the
Euler-Poincar\'e theory.

The results of \cite{HMR[1998a]} allow one to compute
the Euler-Poincar\'e equation for the Lagrangian $\langle{L}\rangle$ in
(\ref{Lag-Eul-B}) depending on the Eulerian mean velocity ${\bf u}$,
and the advected quantities $D$ and $\langle\zeta^k\zeta^l\rangle$ as
\begin{eqnarray}\label{EPeqn-Lbar-E}
0 &=& \left(\frac{\partial}{\partial t}
+ u^j\frac{\partial}{\partial x^j}\right)
\frac{1}{D}\frac{\delta \langle L\rangle}{\delta u^i}
+ \frac{1}{D}\frac{\delta \langle L\rangle}{\delta u^j}u^j_{,i}
\\&&
\qquad
-\
\frac{\partial}{\partial x^i}\,
\frac{\delta \langle L\rangle}{\delta D} \
+\
\frac{1}{D}
\frac{\delta \langle L\rangle}{\delta \langle\zeta^k\zeta^l\rangle}\
\frac{\partial}{\partial x^i}\,
\langle\zeta^k\zeta^l\rangle
\,.
\nonumber
\end{eqnarray}
We compute the following variational derivatives of the averaged approximate
Lagrangian $\langle L\rangle$ in Eq. (\ref{Lag-Eul-B})
\begin{eqnarray}\label{mean-Lag-der-E}
\frac{1}{D} \frac{\delta\langle L\rangle }{\delta {\bf u}}
&=& {\bf u}
 - \frac{1}{D} \Big(\partial_k\,
D\langle\zeta^k\zeta^l\rangle \partial_l\Big) {\bf u}
\equiv {\bf v},
\nonumber\\
\frac{\delta\langle L\rangle }{\delta D}
&=& - P
+ \frac{1}{2}|{\bf u}\,|^2
+ \frac{1}{2}\langle\zeta^k\zeta^l\rangle
\big({\bf u}_{,k}\boldsymbol\cdot{\bf u}_{,l}\big)
\equiv -P_{tot},
\nonumber\\
\frac{\delta\langle L\rangle }{\delta P}
&=& 1 - D \,,
\nonumber\\
\frac{\delta \langle L\rangle}{\delta \langle\zeta^k\zeta^l\rangle}
&=& \frac{D}{2}\big({\bf u}_{,k}
\boldsymbol\cdot{\bf u}_{,l}\big)
\,.
\end{eqnarray}
The Euler-Poincar\'e Eq. (\ref{EPeqn-Lbar-E})
for this averaged Lagrangian takes the form,
\begin{eqnarray}\label{EPeqn-EMM}&&
\frac{\partial{\bf v}}{\partial t}
+ {\bf u}\boldsymbol{\cdot\nabla}{\bf v}
+ v_j \boldsymbol\nabla\,  u^j
+ \boldsymbol\nabla\,P_{tot} =
-\ \frac{1}{2}\,
\big({\bf u}_{,k}\boldsymbol\cdot{\bf u}_{,l}\big)
\boldsymbol\nabla\,
\langle\zeta^k\zeta^l\rangle
\,,
\\&&\hspace{-.85in}
\quad\hbox{where}\quad
{\bf v} = {\bf u} - \tilde\Delta_D{\bf u}
\quad\hbox{with}\quad
\tilde\Delta_D
\equiv
\frac{1}{D}
\big(\partial_k\,D\,
\langle\zeta^k\zeta^l\rangle \partial_l\big)
\quad\hbox{and}\quad
\boldsymbol{\nabla\,\cdot}{\bf u}=0
\,.
\label{vee-redef-E}
\end{eqnarray}
Its definition as a variational derivative indicates that
${\bf v}$ is a specific momentum in a certain sense dual to the
velocity ${\bf u}$. For more discussion of physical interpretations
of ${\bf u}$ and ${\bf v}$, see \cite{Holm[1999]}.
The Euler-Poincar\'e equations (\ref{EPeqn-EMM}) -- (\ref{vee-redef-E}) define
the Eulerian mean motion (EMM) model.
Incompressibility of the Eulerian mean velocity ${\bf u}$ follows from the
continuity equation (\ref{cont-eqn-A}) and the constraint
$\delta\langle{L}\rangle/\delta P = 0$.  A natural set of  boundary conditions 
is 
\begin{equation} \label{bc-E}
{\bf v}\boldsymbol{\cdot\hat{n}}=0\,,
\quad
{\bf u} = 0\,,
\quad\hbox{and}\quad
\boldsymbol{\hat{n}\,\cdot}
\boldsymbol{\langle\xi\xi\rangle}
=0,
\quad\hbox{on a fixed boundary.}
\end{equation}
Then, provided the Helmholtz operator $1-\tilde\Delta_D$ for $D=1$ may be
inverted, the Eulerian mean pressure $P$ may be obtained by solving an elliptic
equation.


\subsection{Reducing the EMM equation to the $n$-dimensional CH equation}

When the Eulerian mean covariance is isotropic and homogeneous, so that
$\langle\zeta^k\zeta^l\rangle=\alpha^2\delta^{kl}$ (for a constant length
scale  $\alpha$, whose magnitude is set by the initial conditions for the
Eulerian mean covariance) then the EMM equation (\ref{EPeqn-EMM}) {\bf
reduces}
to the $n$-dimensional Camassa-Holm equation introduced
in~\cite{HMR[1998a]},~\cite{HMR[1998b]}, namely,
\begin{eqnarray}\label{nd:CHeqn}
\frac{\partial{\bf v}}{\partial t}
+ {\bf u}\boldsymbol{\cdot\nabla}{\bf v}
+ v_j \boldsymbol\nabla  u^j
+ \boldsymbol\nabla\,P_{tot} = 0\,,
\quad \boldsymbol{\nabla\cdot}{\bf u}
=0=
\boldsymbol{\nabla\cdot}{\bf v}\,,
\\
\quad \hbox{where} \quad
{\bf v} = {\bf u} - \alpha^2\Delta {\bf u}\,,
\quad
P_{tot} = P - \frac{1}{2}|{\bf u}\,|^2
- \frac{\alpha^2}{2}|\nabla{\bf u}\,|^2
\,.\label{vee-def-CH}
\end{eqnarray}
This $n$-dimensional CH equation set is an invariant subsystem of the
Euler-Poincar\'e system (\ref{EPeqn-EMM}), with definition (\ref{vee-redef-E})
and advection law (\ref{cov-dyn-E}), because the homogeneous isotropic initial
condition $\langle\zeta^k\zeta^l\rangle = \alpha^2\delta^{kl}$ is invariant
under
the dynamics of equation (\ref{cov-dyn-E}). Hence, any of the formulae above
remain valid if we set $\langle\zeta^k\zeta^l\rangle =
\alpha^2\delta^{kl}$, with constant $\alpha$.

Equations of the type (\ref{nd:CHeqn}) but with additional
dissipative terms were considered previously in the theory of second
grade fluids~\cite{DF1974} and were treated recently in the
mathematical literature~\cite{CV1996},~\cite{CV1997}. Second grade
fluid models are derived from continuum mechanical principles of
objectivity and material frame indifference, after which thermodynamic
principles such as the Clausius-Duhem relation and stability of
stationary equilibrium states are imposed to restrict the allowed
values of the parameters in these models. In contrast, the CH equation
(\ref{nd:CHeqn}) is derived here by applying asymptotic expansions,
Eulerian means, and an assumption of isotropy of fluctuations in
Hamilton's principle for an ideal incompressible fluid. This derivation
provides the interpretation of the length scale $\alpha$ as the typical
amplitude of the rapid fluctuations whose Eulerian mean is taken in Hamilton's
principle.

The $n$-dimensional CH equation (\ref{nd:CHeqn}) implies the conservation
of energy $\frac{1}{2}\int{d^{\,3}x}\,{\bf u}\,
\boldsymbol\cdot\,{\bf v}$ and helicity
$\frac{1}{2}\int{d^{\,3}x}\,{\bf v}\,
\boldsymbol\cdot\,{\rm curl}{\bf v}$. Its
steady vortical flows include the analogs of the Beltrami flows ${\rm
curl}{\bf v}= \lambda {\bf u}$. In the periodic case, we define
${\bf v}_{\bf k}$ as the ${\bf k}$-th Fourier mode of the
specific momentum ${\bf v}\equiv(1-\alpha^2\Delta){\bf u}$;
so that ${\bf v}_{\bf k} \equiv (1+\alpha^2
|{\bf k}|^2){\bf u}_{\bf k}$. Then Eq. (\ref{nd:CHeqn})
becomes~\cite{HMR[1998a]},~\cite{HMR[1998b]}
\begin{equation} \label{CH-spectral}
\Pi_{\perp}\left(
\frac{d}{dt} {\bf v}_{\bf k}
- i \sum_{{\bf p}+{\bf n}={\bf k}}
\frac{ {\bf v}_{\bf p} }{1+\alpha^2|{\bf p}|^2 }
\times ( {\bf n}\times{\bf v}_{\bf n}) \right)
=0,
\end{equation}
where $\Pi_{\perp}$ is the Leray projection onto Fourier modes
transverse to ${\bf k}$ (this ensures incompressibility). Hence,
the nonlinear coupling among the modes is suppressed by the denominator
when $1+\alpha^2|{\bf p}|^2\gg|{\bf n}|$.

An essential feature of the $n$-dimensional CH equation (\ref{nd:CHeqn})
is that its specific momentum ${\bf v}$ is transported by a velocity
${\bf u}$ that is smoothed, or filtered, by application of the inverse elliptic
Helmholtz operator $(1-\alpha^2\Delta)$. The effect on length
scales smaller than $\alpha$ is that steep gradients of the specific
momentum ${\bf v}$ tend not to steepen much further, and that thin
vortex tubes tend not to get much thinner as they are transported.
Furthermore, as our present numerical simulations shall verify, the effect on
length scales larger than $\alpha$ is negligible. Hence,
the $n$-dimensional CH equation preserves the assumptions under which it is
derived.


\subsection{Physical interpretation of ${\bf v}$ as the Lagrangian mean
velocity}

The Stokes mean drift velocity is defined
by~\cite{Andrews-McIntyre[1978a]},
\begin{equation} \label{Stokes-def}
\langle{\bf U}\rangle^S
\equiv
\langle\boldsymbol{\zeta\cdot\nabla}
{\mathbf{u}}^{\,\prime}\,\rangle
\,.
\end{equation}
Hence, Eq. (\ref{u-prime-zeta}) implies
\begin{equation} \label{Stokes-for-EMM}
\langle{\mathbf{U}}\rangle^S
=
-\,
\langle\boldsymbol{\zeta\cdot\nabla}
\boldsymbol{\zeta\cdot\nabla}\rangle {\mathbf{u}}
=
-\,
\tilde\Delta{\mathbf{u}}
+ o(|\boldsymbol\zeta|^2)
\,,
\end{equation}
where
\begin{equation} \label{delta-E-def}
\tilde\Delta
\equiv
\big(\partial_k\,\langle\zeta^k\zeta^l\rangle \partial_l\big)
=
\tilde\Delta_D\big|_{D=1}
\,,
\end{equation}
and we argue that $\boldsymbol{\nabla\cdot\zeta} =
o(|\boldsymbol\zeta|^2)$. Thus, we find that ${\mathbf{v}}$ satisfies, to order
$o(|\boldsymbol\zeta|^2)$,
\begin{equation} \label{V is U-Lag for EMM}
{\mathbf{v}}
\equiv
{\mathbf{u}}
-
\tilde\Delta{\mathbf{u}}
=
{\mathbf{u}}
+
\langle{\mathbf{U}}\rangle^S
=
\langle{\mathbf{U}}\rangle^L
\,.
\end{equation}
Therefore to this order, ${\mathbf{v}}$ in the EMM theory is the Lagrangian
mean velocity.


\subsection{Kelvin circulation theorem for EMM and CH equations}

Being Euler--Poincar\'e, the Eulerian mean motion (EMM) equation
(\ref{EPeqn-EMM}) and its invariant reduced form the CH equation
(\ref{nd:CHeqn}) for $\langle\zeta^k\zeta^l\rangle = \alpha^2\delta^{kl}$ has a
corresponding Kelvin-Noether circulation theorem,
\begin{equation}\label{KelThm-E}
\frac{ d}{dt}\oint_{\gamma({\mathbf{u}})}{\mathbf{v}}
\boldsymbol{\cdot}d{\mathbf{x}}
=
-\,\frac{1}{2}\int\int_{S({\mathbf{u}})}
\boldsymbol{\nabla}
\big({\mathbf{u}}_{,k}\boldsymbol\cdot{\mathbf{u}}_{,l}\big)
\times
\boldsymbol{\nabla}
\langle\zeta^k\zeta^l\rangle
\boldsymbol{\cdot}
d{\mathbf{S}}
\,,
\end{equation}
for any closed curve ${\gamma({\mathbf{u}})}$ that moves with the
Eulerian mean fluid velocity ${\mathbf{u}}$ and surface $S({\mathbf{u}})$ with
boundary ${\gamma({\mathbf{u}})}$. Thus in this Kelvin-Noether
circulation theorem, the presence of spatial gradients in the Eulerian mean
fluctuation covariance $\langle\zeta^k\zeta^l\rangle$ creates
circulation
of the Lagrangian mean velocity ${\mathbf{v}}={\mathbf{u}}  -
\tilde\Delta{\mathbf{u}}$.


\subsection{Vortex stretching equation for the Eulerian mean model}

In three dimensions, the EMM equation (\ref{EPeqn-EMM}) may be expressed
in its equivalent ``curl'' form, as
\begin{equation}\label{EPeqn-EMM-curl}
\frac{\partial }{\partial t}{\mathbf{v}}
- {\mathbf{u}} \times \Big(\boldsymbol{\nabla} \times {\mathbf{v}}\Big)
+ \boldsymbol{\nabla}
\big(P_{tot}
+ {\mathbf{u}}\boldsymbol{\cdot}{\mathbf{v}}\big)
=
-\ \frac{1}{2}\,
\big({\mathbf{u}}_{,k}\boldsymbol\cdot{\mathbf{u}}_{,l}\big)
\boldsymbol{\nabla}
\langle\zeta^k\zeta^l\rangle\,,
\quad \boldsymbol{\nabla\cdot{\mathbf{u}}}=0\,.
\end{equation}
The curl of this equation in turn yields an equation for transport and creation
for the Lagrangian mean vorticity, ${\mathbf{q}}\equiv{\rm
curl}\,{\mathbf{v}}$,
\begin{equation} \label{vortex-stretching-E}
\frac{\partial{\mathbf{q}}}{\partial t}
+ \boldsymbol{{\mathbf{u}}\cdot\nabla{\mathbf{q}}}
= \boldsymbol{{\mathbf{q}}\cdot\nabla{\mathbf{u}}}
-\ \frac{1}{2}\,
\boldsymbol{\nabla}
\big({\mathbf{u}}_{,k}\boldsymbol\cdot{\mathbf{u}}_{,l}\big)
\times
\boldsymbol{\nabla}
\langle\zeta^k\zeta^l\rangle
\,,
\quad \hbox{where} \quad
{\mathbf{q}}\equiv{\rm curl}\,{\mathbf{v}}\,,
\end{equation}
and we have used incompressibility of ${\mathbf{u}}$. Thus
${\mathbf{u}}$ is the transport velocity for the generalized vorticity
${\mathbf{q}}$, and the expected vortex stretching term
${\mathbf{q}}\boldsymbol{\cdot\nabla}{\mathbf{u}}$ is
accompanied by an additional vortex creation term proportional to the
Eulerian mean covariance gradient. Of course, this additional term is also
responsible for the creation of circulation of
${\mathbf{v}}$ in the Kelvin-Noether circulation theorem (\ref{KelThm-E}) and
vanishes when the Eulerian mean covariance is homogeneous in space, thereby
recovering the corresponding result for the three dimensional CH
equation~\cite{HMR[1998a]},~\cite{HMR[1998b]}.


\subsection{Energetics of the Eulerian mean model}

Noether's theorem guarantees conservation of energy for the
Euler-Poincar\'e equations (\ref{EPeqn-EMM}), since the Eulerian mean
Lagrangian $\langle L\rangle$ in Eq. (\ref{Lag-Eul-B}) has no explicit
dependence on time. This constant energy is given by
\begin{equation}\label{cons-erg-def-E}
E_t =  \frac{1}{2}\int d^{\,3}x \Big(|{\mathbf{u}}\,|^2
+ \langle\zeta^k\zeta^l\rangle
{\mathbf{u}}_{,k}\boldsymbol\cdot{\mathbf{u}}_{,l}\Big)
= \frac{1}{2}\int d^{\,3}x\ {\mathbf{u}}\boldsymbol\cdot{\mathbf{v}}
\,.
\end{equation}
Thus, the total kinetic energy is the integrated product of the Eulerian mean
and Lagrangian mean velocities. In this kinetic energy, the Eulerian mean
covariance of the fluctuations couples to the gradients of the Eulerian mean
velocity. So there is a cost in kinetic energy for the system either to
increase
these gradients, or to increase the Eulerian mean covariance.


\subsection{Momentum conservation -- stress tensor formulation}

Noether's theorem also guarantees conservation of momentum for the
Euler-Poincar\'e equation (\ref{EPeqn-EMM}), since the Eulerian mean
Lagrangian $\langle L\rangle$ in Eq. (\ref{Lag-Eul-B}) has no explicit
spatial dependence. In  momentum conservation form,
Eq. (\ref{EPeqn-EMM}) becomes
\begin{equation} \label{EPeqn-EMM2}
\frac{\partial v_i}{\partial t}
= -\ \frac{\partial }{\partial x^j}
\Big(v_i\, u^j + P\delta^j_i
-
{\mathbf{u}}_{,k}\boldsymbol\cdot{\mathbf{u}}_{,i}\,
\langle\zeta^k\zeta^j\rangle
\Big)\,.
\end{equation}
The boundary conditions are given in Eq. (\ref{bc-E}).


\subsection{A second moment turbulence closure model for EMM}

When dissipation and forcing are added to the EMM motion equation
(\ref{EPeqn-EMM}) by using the phenomenological viscosity
$\nu\tilde\Delta{\mathbf{v}}$ and forcing ${\mathbf{F}}$, one finds a {\bf
second moment Eulerian mean turbulence model} given by
\begin{eqnarray}\label{EMM-2pt-eqns}
\Big(\frac{\partial}{\partial t}
+ {\mathbf{u}}\boldsymbol{\,\cdot\nabla}\Big){\mathbf{v}}
+ v_j \boldsymbol\nabla u^j
&+& \boldsymbol\nabla P_{tot}
+\ \frac{1}{2}\,
\big({\mathbf{u}}_{,k}\boldsymbol\cdot{\mathbf{u}}_{,l}\big)
\boldsymbol\nabla
\langle\zeta^k\zeta^l\rangle
\\
&=& \nu\,\tilde\Delta{\mathbf{v}} + {\mathbf{F}}\,,
\quad\hbox{where}\quad
\boldsymbol{\nabla\,\cdot}{\mathbf{u}}=0\,,
\nonumber
\end{eqnarray}
with viscous boundary conditions ${\mathbf{v}}=0$, ${\mathbf{u}}=0$ at a fixed
boundary. Note that the Eulerian mean fluctuation covariance
$\langle\zeta^k\zeta^j\rangle$ appears in the dissipation operator
$\tilde\Delta$. In the absence of the forcing ${\mathbf{F}}$, 
this viscous EMM turbulence model dissipates
the energy $E$ in Eq. (\ref{cons-erg-def-E}) according to
\begin{equation}\label{EMM-erg-dissip}
\frac{dE}{dt}
= -\,\nu \int d^{\,3}x
\Big[ {\rm tr}(\boldsymbol{\nabla}{\mathbf{u}}^T
\boldsymbol{\cdot\,\langle\zeta\zeta\rangle}
\boldsymbol{\,\cdot\nabla}{\mathbf{u}}) +
\tilde\Delta{\mathbf{u}}
\boldsymbol{\cdot}
\tilde\Delta{\mathbf{u}}\,\Big]
\,.
\end{equation}
This negative definite energy dissipation law is a consequence of adding viscosity
with $\tilde\Delta$, instead of using the ordinary Laplacian operator.
In the isotropic homogeneous case  of this model, where
$\langle\zeta^k\zeta^l\rangle=\alpha^2\delta^{kl}$ (for a constant length
scale
$\alpha$) we find the viscous Camassa-Holm equation (VCHE), or the
Navier-Stokes alpha model,
\begin{equation}\label{VCHE-eqns}
\Big(\frac{\partial}{\partial t}
+ {\mathbf{u}}\boldsymbol{\,\cdot\nabla}\Big){\mathbf{v}}
+ v_j \boldsymbol\nabla u^j
+ \boldsymbol\nabla P_{tot}
= \nu\alpha^2\,\Delta{\mathbf{v}} + {\mathbf{F}}\,,
\quad\hbox{where}\quad
\boldsymbol{\nabla\,\cdot}{\mathbf{u}}=0\,,
\end{equation}
where $\Delta$ is the usual Laplacian operator for this case and
${\mathbf{v}}$ and
$P_{tot}$ are defined in (\ref{vee-def-CH}).


\subsection{Constitutive interpretation of the VCHE}

Chen et al.~\cite{Chen etal[1998a]}--~\cite{Chen
etal[1998c]} gave a continuum mechanical interpretation to the VCHE
closure model, by rewriting the VCHE (\ref{VCHE-eqns}) in the equivalent {\bf
constitutive form},
\begin{equation}\label{constit-eqn}
\frac{d{\mathbf{u}}}{dt} = {\rm
div}\hbox{\bf T}\;,\;\hbox{\bf T} =  -p\hbox{\bf I} + 2\nu (1 -
\alpha ^2\Delta)\hbox{\bf D} + 2\alpha ^2 {\bf\dot{\hbox{\bf D}}}\;,
\end{equation}
with $\boldsymbol\nabla\cdot{\mathbf{u}}=0, \hbox{\bf D} = (1/2)
(\boldsymbol\nabla{\mathbf{u}} + \boldsymbol\nabla{\mathbf{u}}^T),\
{\boldsymbol{\Omega}} =
(1/2) (\boldsymbol\nabla{\mathbf{u}} - \boldsymbol\nabla{\mathbf{u}}^T)$, and
co-rotational (Jaumann) derivative given by ${\bf\dot{\hbox{\bf D}}} =
d\hbox{\bf D}/dt  + \hbox{\bf D}\,{\boldsymbol{\Omega}} -
{\boldsymbol{\Omega}} \hbox{\bf D}$, with $d/dt =
\partial/\partial t + {\mathbf{u}}\boldsymbol{\cdot\nabla}$.  In this
form, one recognizes the constitutive form of VCHE as a
variant of the rate-dependent incompressible homogeneous fluid of second
grade~\cite{Dunn-Fosdick[1974]},~\cite{Dunn-Rajagopal[1995]}, whose
viscous dissipation, however, is {\it modified} by the Helmholtz operator
$(1 - \alpha^2\Delta)$.  There is a tradition at least since
Rivlin~\cite{Rivlin[1957]} of modeling turbulence by using continuum
mechanics principles such as objectivity and material frame indifference
(see also~\cite{Chorin[1988]}).  For example, this sort of approach is
taken in deriving Reynolds stress algebraic equation
models~\cite{Shih-Zhu-Lumley[1995]}. Rate-dependent closure models of
mean turbulence such as the VCHE have also been obtained by a two-scale
DIA approach~\cite{Yoshizawa[1984]} and by renormalization group
methods~\cite{Rubinstein-Barton[1990]}.


\subsection{Comparison of VCHE with LES and RANS models}

Reynolds-averaged Navier-Stokes (RANS) models of turbulence are part of the
classic theoretical development of the
subject~\cite{Hinze[1975]},~\cite{Townsend[1967]},~\cite{Lumley-Tennekes[1972]}.
The related Large Eddy Simulation (LES) turbulence modeling
approach~\cite{WReynolds[1987]},~\cite{Piomelli[1993]},~\cite{L&M[1996]},
provides an operational definition of the intuitive idea of Eulerian resolved
scales of motion in turbulent flow. In this approach a filtering function
${\cal
F}({\mathbf{r}})$ is introduced and the Eulerian velocity field ${\mathbf{U}}_E$ is
filtered in an integral sense, as
\begin{equation} \label{u-fltrd}
\bar{{\mathbf{u}}}({\mathbf{r}})\equiv \int_{{R}^3} d^3{r}^{\,\prime}\,
   \,{\cal F}({\mathbf{r}}-{\mathbf{r}}^{\,\prime})
   \,{\mathbf{U}}_E\,({\mathbf{r}}^{\,\prime})\,.
\end{equation}
This convolution of ${\mathbf{U}}_E$ with ${\cal F}$ defines the large scale,
resolved, or filtered velocity, $\bar{\mathbf{u}}$. The corresponding small
scale, or subgrid scale velocity,
${\mathbf{u}}^{\,\prime}$, is then defined as the difference,
\begin{equation}\label{u-small}
 {\mathbf{u}}^{\,\prime}({\mathbf{r}})\equiv
{\mathbf{U}}_E\,(\mathbf{r})-\bar{\mathbf{u}}\,(\mathbf{r})\,.
\end{equation}
When this filtering operation is applied to the Navier-Stokes system, the
following dynamical equation is obtained for the filtered velocity,
$\bar{\mathbf{u}}$, cf. Eq. (\ref{constit-eqn}),
\begin{equation} \label{NS-fltrd}
    \frac{\partial}{\partial t}\bar{{\mathbf{u}}}
   + \bar{{\mathbf{u}}}\boldsymbol{\cdot\nabla}\bar{{\mathbf{u}}}
 = -\,{\rm div}\overline{\hbox{{\bf T}}}
   -\,\boldsymbol\nabla
   \bar{p}+\nu\,\Delta \bar{{\mathbf{u}}}\,,
   \quad \boldsymbol{\nabla\cdot\bar{{\mathbf{u}}}}=0\,,
\end{equation}
in which $\bar{p}$ is the filtered pressure field (required to maintain
$\boldsymbol{\nabla\cdot\bar{\mathbf{u}}}=0$) and the tensor difference
\begin{equation} \label{stress-dev}
  \overline{\hbox{{\bf T}}} =\overline{
({\mathbf{U}}_E{\mathbf{U}}_E)}-\bar{\mathbf{u}}\bar{\mathbf{u}}\,,
\end{equation}
represents the subgrid scale stress due to the turbulent eddies.  This subgrid
scale stress tensor appears in the same form as the Reynolds stress tensor
obtained from Reynolds averaging the Navier-Stokes equation.

The results of Chen et al.~\cite{Chen etal[1998a]}--~\cite{Chen etal[1998c]},
may be given either an LES, or RANS interpretation simply by comparing the
constitutive form of the VCHE in (\ref{constit-eqn}) term by term with equation
(\ref{NS-fltrd}), provided one may ignore the difference between Eulerian mean,
and Lagrangian mean velocities as being of higher order. Additional LES
interpretations, discussions and numerical results for forced-turbulence
simulations of the VCHE model, or the Navier-Stokes alpha model,
 will be presented  below.


\section{Direct Numerical Simulation, Related Definitions and Energy Spectra}

In Fourier space, the viscous alpha model equation (\ref{EPeqn-EMM-curl})
 with isotropic viscosity can be written for the Lagrangian mean velocity 
${\bf v}$ as follows:

\begin{equation}
\label{ns-k}
\frac{\partial {\bf v}_{\bf k}}{\partial t} = 
{\hat P}({\bf k}) ({\bf u} \times 
{\bf q})_{\bf k} - \nu k^2 {\bf v}_{\bf k} + {\bf f}_{\bf k}, 
\end{equation}
\begin{equation}
\label{in-k}
{\bf k}\cdot {\bf v}_{\bf k} = 0,
\end{equation}
where ${\hat P}_{ij} = \delta_{ij} - k_ik_j/k^2$ is the incompressible
projection operator. The Eulerian mean velocity satisfies
 ${\bf u} = (1 - \alpha^2 \Delta)^{-1} {\bf v}$
(or ${\bf u}({\bf k}) = (1 + \alpha^2 k^2)^{-1} {\bf v}({\bf k})$),
${\bf q} = \nabla \times {\bf v}$ and ${\bf f}_{\bf k}$ is the forcing term
for the ${\bf k}$-th velocity component.
We emphasize that $1/\alpha$ acts as the cutoff wavenumber for the
nonlinearity in alpha model. 

A pseudo-spectral code has been developed
for solving the above equations
in a cubic box for periodic boundary conditions. The code is written in
message-passing language for the SGI-ORIGIN-2000 machine in the Advanced
Computing Laboratory at the Los Alamos National Laboratory.
The viscous term is integrated analytically in time.
The other terms are discretized using a second-order Adams-Bashforth scheme.
The nonlinear terms are calculated using pseudo-spectral
methods\cite{chen-shan}. The time evolution of Eq. (\ref{ns-k}) can be written:

\begin{eqnarray}
\frac{{\bf v}_{\bf k}^{n+1} - 
{\bf v}_{\bf k}^{n}exp(-\nu k^2\Delta t)}{\Delta t} & = & {\hat P}({\bf k}) 
[\frac{3}{2}({\bf u}\times{\bf q})_{\bf k}^{n}exp(-\nu k^2\Delta t)
\hspace{2.in}\nonumber \\
&-&\frac{1}{2}({\bf v}\times{\bf q})_{\bf k}^{n-1}
exp(-2\nu k^2\Delta t)] + f_{\bf k}. 
\end{eqnarray}

In this paper, we adopt the following definitions:
the mean velocity fluctuation, $u'$, is defined as
\[
u'=\sqrt{\frac{2}{3}E_t}=(\frac{2}{3} \int_o^{\infty} E(k)dk )^{\frac{1}{2}};
\]
where $E_t$ is the total energy defined in Eq. (\ref{cons-erg-def-E}) and
 $E(k)$ is the energy spectrum:
\[ E(k) = \sum_{{\bf k}-1/2}^{{\bf k}+1/2} {\bf u}({\bf k}')\cdot{\bf v}({\bf k}').\]
The Taylor microscale and the mean dissipation rate are defined respectively
as follows:
\[
\lambda=(15\nu/\epsilon)^{1/2} u',
\]
\[
\epsilon=2\nu \int_o^{\infty}k^2E(k)dk.
\]
The large eddy turn-over time is given by:
\[
\tau = \frac{L_f}{u'}. 
\]
where $L_f$ is the integral length. 
The Kolmogorov dissipation scale $\eta$ is defined as
$({\nu^3}/{\epsilon})^{\frac{1}{4}}$, 
with corresponding wave number $k_d={1}/{\eta}$.
The Taylor microscale Reynolds number is defined by
\[
R_{\lambda}=\frac{u'\lambda}{\nu}.
\]

To maintain a statistical steady state, the forcing term ${\bf f}_{\bf k}$
 was introduced in the
first two shells of the Fourier modes ($k<2.5$), in which the kinetic energy
of each mode in the two shells was forced to be
constant in time, while the energy ratio between shells consistent with
$k^{-5/3}$ in order to approximate a larger inertial sub-range\cite{chen-shan}.  
Most simulations were carried out for about ten large eddy
turnover times before recording any data. The initial velocity
was a Gaussian field with a prescribed energy spectrum:
 $\sim k^4 exp[- (k/k_0)^2]$, where
$k_0$ is a constant whose value is approximately $5$. Statistics were
obtained by averaging physical quantities over several eddy turnover
times.

DNS has been carried out for three cases: with $\alpha = 0$ (the Navier-Stokes
equations), $1/32$ and $1/8$ for the same viscosity $\nu = 0.001$.
The corresponding $R_{\lambda}$ are $147, 182$ and $279$, respectively. 
In Fig.~1, we show the energy spectra of the three simulations.  We note that
because each simulation used the same viscosity, the higher Reynolds number 
flows correspond to smaller Taylor 
microscales, leading to more compact energy spectra. This is similar to
the results of subgrid simulations for higher Reynolds number flows\cite{wang}.
In Fig.~2, we compare the energy spectrum for $\alpha = 1/8$ with 
mesh sizes of $256^3$ and $64^3$. It is seen that the energy 
spectrum at the large scales (in the inertial range) are the same, 
indicating that the large scale flow properties for $\alpha = 1/8$
can be preserved in a less-resolved simulation. 

To see how the energy cascades from the large scale modes to the small scale
modes, in Fig.~3 we show the energy transfer spectrum, 
\[ \Pi (k) \equiv {\bf v}({\bf k})\cdot {\hat P}({\bf k}) ({\bf u} \times
{\bf q})_{\bf k}, \]
as a function of the wave number, $k$. We note that the energy transfer 
spectrum for all three cases with different $\alpha$'s are constant and
agree quite well in the inertial range for $k < 20$, except for the effect
of noise. This result indicates that the alpha model preserves the 
fundamental properties of the Kolmogorov energy cascade in the inertial 
range\cite{wang}. For the dissipation range ($k > 20$), however, the change of
the energy transfer spectrum is significant for $\alpha = 1/8$. 

\section{Vorticity Structures and Alignment}

As contrasted with the hyperviscosity approach\cite{borue,ccs} in which the normal
dissipation operator is replaced by a higher order Laplace operator and therefore
only small scale fluid motions are affected,  in the alpha model the nonlinear
vortex stretching dynamics is modified as shown in Eq. (\ref{vortex-stretching-E}),
where the vorticity ${\bf q}$ is defined
by the curl of the velocity ${\bf v}$ while the advective velocity is
 ${\bf u} = (1 - \alpha^2 \Delta)^{-1} {\bf v}$. For $\alpha$ not zero, 
using a smoothed velocity ${\bf u}$ rather than the original velocity ${\bf v}$
suppresses the vortex stretching dynamics, especially for the small scale vortices 
whose size is less than $\alpha$.

To give a qualitative idea about how the vortex structures change with increasing
$\alpha$, in Fig.~4 (a-c) we present the iso-surfaces of vorticity 
for $q / q' = 2$ when $\alpha = 0, 1/32$ and $1/8$, where
$q' = \langle q^2 \rangle^{1/2}$ is the {\em root-mean-square} value of 
${\bf q}$. It is seen that the tube-like vortex structures\cite{cao-chen} persist
in all three simulations, implying that the alpha model does not change
the qualitative feature of stretching physics. On the other hand, it is evident
that with increasing of $\alpha$, the vortex aspect ratio (characteristic vortex
radius/characteristic vortex length) decreases. A similar phenomenon has been noticed
in hyperviscosity simulations also\cite{borue,ccs} (where the vortex 
stretching dynamics is not suppressed) and in subgrid model simulations\cite{wang}.
However, the physical mechanisms for this phenomena are rather 
different in the alpha model 
from the actions of hyperviscosity, since the alpha term affects the nonlinearity
(the cause of vortex stretching), while hyperviscosity does not.

One of the most important properties of the vortex stretching dynamics in
the Navier-Stokes turbulence is the so-called alignment phenomenon\cite{ashurst},
in which the three-dimensional vorticity is locally preferentially aligned with the 
direction of the second eigenvector of the symmetric strain-rate tensor,
$S_{ij} = 1/2 (\partial u_i/\partial x_j + \partial u_j/\partial x_i)$. In
three-dimensional space, $S_{ij}$ is a $3\times3$ matrix and
has three eigenvalues, $\lambda_1, \lambda_2$ and $\lambda_3$. The 
incompressibility condition leads to: 
\[ \lambda_1 + \lambda_2 + \lambda_3 = 0. \]
Assume $\lambda_1 \ge \lambda_2 \ge \lambda_3 $, then $\lambda_1 \ge 0$
and $\lambda_3 \le 0$. For the Navier-Stokes turbulence, it has been 
found\cite{ashurst} that the spatially averaged $\lambda_2$ is greater than zero. 
Therefore in Navier-Stokes turbulence, the stretching dynamics of vorticity  
is dominated by two directional expansions with positive eigenvalues and  
one directional contraction with negative eigenvalue.  
Since the alpha model is equivalent to filtering the transport
velocity in the stretching term for the vorticity equation 
Eq. (\ref{vortex-stretching-E}), it is interesting to 
study how the alignment phenomenon changes in the alpha model when
compared with the Navier-Stokes model. 

In Fig.~5 (a-c), we present the probability density functions (PDFs) of the
cosine of the angles between the local vorticity and the
eigenvectors of the local strain-rate tensor for $\alpha = 0,
1/32$ and $1/8$. The solid lines, the dotted lines and 
the dotted-dash lines correspond to the maximum eigenvalue, the 
middle eigenvalue and the minimum eigenvalue, respectively. As expected,
the results in Fig.~5(a) (the Navier-Stokes case) are very similar to those
results in \cite{ashurst}. The major feature in this plot is that the
PDF corresponding to the middle eigenvalue $cos(\theta_2)$ is peaked when 
$cos(\theta_2) = \pm 1$, implying the alignment mentioned earlier. 
The PDF of $cos(\theta_3)$ is peaked at the origin, indicating
that the vorticity is locally perpendicular to the direction
associated with the minimum eigenvalue.  
The PDF of $cos(\theta_1)$ is almost flat, implying that the direction
of vorticity is essentially decorrelated from the direction of the 
maximum eigenvalue direction. With increasing $\alpha$, the property 
of  the minimum eigenvalue direction is qualitatively the same, but the PDF
of $cos(\theta_1)$ starts forming a peak at $\pm 1$. For $\alpha = 1/8$, the
PDF values for $cos(\theta_1)$ at $\pm 1$ are even bigger than those of 
$cos(\theta_2)$. This result is quite different from the case of the Navier-Stokes 
equation.  We suspect that this new alignment phenomenon (that the vorticity
aligns with the direction of the maximum eigenvalue) is connected with the 
observation (as shown in Fig~4), that the high amplitude vorticity gets 
thicker as $\alpha$ increases, a result that has also been observed 
in other subgrid simulations\cite{wang}.

To further quantify the change of eigenvalues as a function of $\alpha$, 
in Fig.~6(a-c), we show the PDFs of the eigenfunctions. 
There are two essential features in these plots: (i) as $\alpha$ increases,
$\langle \lambda_2 \rangle$ keeps positivity, consistent with 
the result in the Navier-Stokes turbulence; (ii) all eigenvalues
tend toward smaller values as $\alpha$ increases, meaning 
that the stretching is being suppressed in the alpha model. To study the mean effect
of the stretching term on the growth of the enstrophy, 
$\langle{\bf q}\cdot{\bf q}\rangle$,
in Fig.~7 we present the PDF of $ Z \equiv {\bf q}\cdot{\hat S}\cdot{\bf q}$ 
as a function of $\alpha$. The PDFs of these three quantities show very strong
intermittency. The flatnesses are 
$F_{q_i S_{ij}q_j} = 1965 $ for $\alpha = 0$, $3566$ for $\alpha = 1/32$
and $5319$ for $\alpha = 1/8$. The PDFs are strongly positively skewed
($24.28, 35.71$ and $47.02$) which is consistent with the expectation that
the stretching term in the vorticity equation gives a net
positive contribution to the dynamics of enstrophy. 

\section{Two-Point Statistics}

Two-point statistics, primarily the scaling relations of the velocity structure 
functions in the inertial range, were proposed by Kolmogorov in 1941\cite{k41} 
based on the self-similarity hypothesis and in 1962\cite{k62} using the idea of 
the refined similarity hypothesis. Understanding the inertial range intermittent 
dynamics of fluid turbulence has been a very important focal point for
the last few decades, and the two-point inertial range statistics have 
been extensively examined in experiments at high Reynolds numbers and by 
numerical simulations at low to moderate Reynolds numbers\cite{ccs}. 
In this section, we study the probability density function of the
velocity increment and the scalings of the velocity structure functions
for systems with various values of  $\alpha$ using DNS data, and then
compare our measurements with the Kolmogorov theories.

In Fig.~8, we compare the PDFs of the longitudinal velocity gradient,
$\partial u/\partial x$, for various alpha values. The derivative
here is calculated by a central difference in physical space and
a spatial averaging is used for the ensemble averaging in
the normalization factor calculation. The main 
information in this plot is that the PDFs for all three alpha values 
have close to an exponential tail for large amplitude events. However, the flatnesses
of the PDFs (5.0, 4.1 and 3.8 for $\alpha = 0, 1/32$ and $1/8$,  
respectively) decrease with increasing $\alpha$. The observed reduction of 
small scale intermittency is consistent with the energy spectra in Fig.~1 
where the high $k$ energy spectra are truncated 
for large $\alpha$ values. The skewnesses of the PDFs are $-0.48, -0.51$ and
$-0.54$. In Fig.~9, we present the normalized PDFs of the velocity increment,
\[ \Delta_r u \equiv u(x+r) - u(x) \]
in the inertial range  when the separation $r = 20$ (in mesh units). 
It is seen that all three PDFs collapse quite well for most values of 
$\Delta_r u$, implying that the fundamental
features of the two-point statistics in the inertial range do not change
much as $\alpha$ varies. This is a very desirable feature in using the alpha
model as a subgrid model.

In Fig.~10, we show the second-order structure function, 
$\langle \Delta_r u\rangle$, versus $r$ at
different alpha values. It is seen that when $r$ less than $5$ mesh points,
the Taylor expansion gives a simple scaling: $\langle \Delta_r u\rangle 
\sim r^2$. In the narrow inertial range ($20 < r < 60$), the simulation
results with $\alpha = 0$ and $1/32$ agree qualitatively with the
Kolmogorov $2/3$ scaling\cite{k41}. The slightly slower than $2/3$ growth may
indicate the intermittent correction. For the case of $\alpha = 1/8$, 
the filtering of small scale motions significantly decreases the structure 
function in the inertial range. For this case, the $2/3$ scaling can only 
been seen for the very narrow region $40 < r < 60$. 

In Fig.~11, we present the flatness of the velocity increment,
$\langle (\Delta_r u)^4 \rangle/langle (\Delta_r u)^2 \rangle^2$,
versus the normalized separation, $r/r_0$. The normalization length is
taken to be half the box length $\pi$. As expected, in 
the small scale region with $r/r_0 < 0.005$, the alpha model is quite 
different from the Navier Stokes dynamics and the Navier-Stokes equation
has the largest flatness value, implying that the Navier-Stokes 
turbulence is more intermittent than the alpha model dynamics, consistent
with the vortex visualization in Fig.~4. It is very interesting to note,
however,  that the 
flatnesses versus $r/r_0$ are almost identical for all three alpha values studied
in the narrow inertial range. The decaying of the flatness as a function
of $r/r_0$ implies the existence of an intermittency correction for the
scaling exponents. The numerical measurement gives 
$\langle (\Delta_r u)^4 \rangle/\langle (\Delta_r u)^2 \rangle^2 \sim
r^{-0.12}$ in the inertial range. This scaling exponent agrees qualitatively
 with the value of $-0.11$ from other direct numerical simulation
results\cite{ccs}. 

\section{Concluding Remarks}

Our main objective in this paper has been to investigate the utility of the 
recently proposed alpha equations as a subgrid scale model for three-
dimensional turbulence.  Our main tool has been direct numerical 
simulation (DNS) of turbulence in a periodic box. Our conclusions are 
based on comparisons of DNS using the Navier-Stokes alpha equations for several 
values of the alpha parameter, including the limiting case ${\alpha = 0}$ in 
which the Navier-Stokes equations are recovered.  Our principal 
conclusion is that the alpha model simulations can reproduce most of the 
large scale features of Navier-Stokes turbulence {\it even when these 
simulations do not resolve the fine scale dynamics}, at least in the case of 
turbulence in a periodic box.

We summarized known analytic properties of the alpha models, including 
an outline of their derivation and the associated assumptions, their 
simplification for the case of constant alpha, and their conservation 
properties.  We also offered interpretations of nonlinear dynamics of the 
alpha models and indicated the changes one might expect from the 
dynamics of the Navier-Stokes equations.

One of our principal computational results is shown in Fig. 3, where two DNS 
simulations at ${\alpha = 1/8}$, but using two different resolutions, are 
compared with each other and with a Navier-Stokes simulation 
(representing truth).  The spectra are identical for the larger wavelengths, 
demonstrating (in this case) that one does not need to resolve the small 
scale dynamics to reproduce the large scale features of the turbulence.

We further compared vorticity structures and alignment, and also two point 
statistics to illustrate the altered dynamics of the alpha models.  In general, 
we found consistency of these DNS results with our expectations based 
on analysis.

The use of the alpha equations as a subgrid model of turbulent flows in 
more complicated geometries and forcings remains to be studied.  
However we believe these first results are very promising.

\section{Acknowledgments}

We thank C. Foias, R. H. Kraichnan, J. C. McWilliam, E. Olson, K. R. Sreenivasan,
 E.S. Titi and S. Wynne, for helpful discussions.
Numerical simulations were carried out at the Advanced Computing Laboratory at 
Los Alamos National Laboratory using the SGI machines.
We appreciate the support and computer time provided by the
US Department of Energy Climate Change Prediction Program.

\section{Figure Captions}
\begin{description}

\item{Fig.~1.}
The energy spectrum, $E(k)$, versus the wave number $k$ for
$\alpha = 0$ (solid line), $1/32$ (dotted line) and $1/8$ (dotted-dash line).
In the inertial range ($k < 10$), a power spectrum with $k^{-5/3}$ can
be identified.

\item{Fig.~2.}
Energy spectra for $\alpha = 1/8$ with $256^3$ (dotted line)
and $64^3$ (dashed-dot line), and compared with the Navier-Stokes simulation
(the solid line). 

\item{Fig.~3.}
The energy transfer spectrum, $\Pi(k)$ versus $k$ for
$\alpha = 0$ (solid line), $1/32$ (dotted line) and $1/8$ (dotted-dash line).

\item{Fig.~4.}
Three-dimensional view of vorticity iso-surfaces for $\alpha = 0$ (a),
$1/32$ (b) and $1/8$ (c) when $q/q' = 2$, where $q'$ is the
{\em rms} of the vorticity. The data are from $256^3$ simulations and
only one eighth of the simulation domain is shown.

\item{Fig.~5.}
The probability density function of $cos(\theta)$ for 
simulations with $\alpha = 0$ (a), $1/32$ (b) and $1/8$ (c). In each 
plot, the solid line, the dotted line and the dotted-dash line
are for the maximum, middle and minimum eigenvalue, respectively. 

\item{Fig.~6.}
The probability density functions for the maximum eigenvalue
$\lambda_1$ (a), the middle eigenvalue $\lambda_2$ (b) and the minimum
eigenvalue $\lambda_3$ (c). The solid line, the dotted line and the
dotted-dash line are for $\alpha =0, 1/32$ and $1/8$, respectively. 

\item{Fig.~7.}
Normalized PDFs of $ z = q_i S_{ij}q_j$ for 
$\alpha =0$ (solid line), $1/32$ (dotted line) and
$1/8$ (dotted-dash line). 

\item{Fig.~8.}
Normalized PDFs of the longitudinal velocity gradient, $\partial u/\partial x$
for $\alpha =0$ (solid line), $1/32$ (dotted line) and
$1/8$ (dotted-dash line).

\item{Fig.~9.}
Normalized PDFs of the velocity increment in the inertial range for
$\alpha =0$ (solid line), $1/32$ (dotted line) and
$1/8$ (dotted-dash line).

\item{Fig.~10.}
The second order structure function as a function of $r$ for
$\alpha =0$ (solid line), $1/32$ (dotted line) and
$1/8$ (dotted-dash line). The dashed line is the scaling prediction 
by Kolmogorov\cite{k41}.

\item{Fig.~11.}
The flatness, $\langle (\Delta_r u)^4 \rangle/\langle (\Delta_r u)^2 \rangle^2$, 
as a function of separation $r/r_0$ for $\alpha =0$ (solid line), 
$1/32$ (dotted line) and $1/8$ (dotted-dash line). Here $r_0$ is taken as 
half the box size, $\pi$.

\end{description}


\begin{thebibliography}{unsrt}

\bibitem{HMR[1998a]}  D.D. Holm, J.E. Marsden, T.S. Ratiu,
Adv. in Math. {\bf 137} (1998) 1.

\bibitem{HMR[1998b]}  D.D. Holm, J.E. Marsden, T.S. Ratiu,
Phys. Rev. Lett. {\bf 80} (1998) 4173.

\bibitem{CH[1993]} R. Camassa and D.D. Holm,
Phys. Rev. Lett. {\bf 71} (1993) 1661.

\bibitem{HKMRS[1998]} D.D. Holm, S. Kouranbaeva, J.E. Marsden,
T. Ratiu and S. Shkoller,
{\it Fields Inst. Comm., Arnold Vol. 2}, Amer. Math. Soc., (Rhode
Island) (1998) to appear.

\bibitem{Holm[1999]} D.D. Holm,
Physica D, to appear.

\bibitem{Chen etal[1998a]} S. Chen, C. Foias, D.D. Holm, E.
Olson, E.S. Titi, S. Wynne,
 Phys. Rev. Lett., {\bf 81} (1998) 5338.

\bibitem{Chen etal[1998b]} S. Chen, C. Foias, D.D. Holm, E.
Olson, E.S. Titi, S. Wynne,
Phys. Fluids, to appear.

\bibitem{Chen etal[1998c]} S. Chen, C. Foias, D.D. Holm, E.
Olson, E.S. Titi, S. Wynne,
Physica D, to appear.

\bibitem{Zagarola96} M.V. Zagarola, Ph.D thesis, Princeton University (1996).

\bibitem{FHT98} C. Foias, D.D. Holm and E.S. Titi,
{\em The 3D viscous Camassa--Holm equations: their relation to the
Navier--Stokes equations and turbulence Theory}, in preparation.

\bibitem{DF1974}
J.E. Dunn and R.L. Fosdick,
Arch. Rat. Mech. Anal., {\bf 56} 191 (1974).

\bibitem{CV1996}
D. Cioranescu and V. Girault,
{C. R. Acad. Sc. Paris} S\'erie 1, {\bf 322} (1996) 1163.

\bibitem{CV1997}
D. Cioranescu and V. Girault,
Int. J. Non-Lin. Mech. {\bf 32} (1997) 317.

\bibitem{Andrews-McIntyre[1978a]} D.G. Andrews and M.E. McIntyre,
J. Fluid Mech. {\bf 89} (1978) 609.

\bibitem{Dunn-Fosdick[1974]} J.E. Dunn and R.L. Fosdick,
Arch. Rat. Mech. Anal. {\bf 56} (1974) 191.

\bibitem{Dunn-Rajagopal[1995]} J.E. Dunn and K.R. Rajagopal,
Int. J. Engng. Sci. {\bf 33} (1995) 689.

\bibitem{Rivlin[1957]} R. S. Rivlin,
Q. Appl. Math. {\bf 15} (1957) 212.

\bibitem{Chorin[1988]} A. J. Chorin,
Phys. Rev. Lett. {\bf 60} (1988) 1947.

\bibitem{Shih-Zhu-Lumley[1995]} T.H. Shih, J. Zhu and J.L. Lumley,
Comput. Methods Appl. Mech. Engng. {\bf 125} (1995) 287.

\bibitem{Yoshizawa[1984]} A. Yoshizawa,
Phys. Fluids, {\bf 27} (1984) 1377.

\bibitem{Rubinstein-Barton[1990]}  R. Rubinstein and J.M. Barton,
Phys. Fluids A, {\bf2} (1990) 1472.

\bibitem{Hinze[1975]} J.O. Hinze, {\it Turbulence}, (Mc-Graw-Hill:
New York, 2nd edition, 1975).

\bibitem{Townsend[1967]}  A.A. Townsend, {\it The Structure of
Turbulent Flows}, (Cambridge University Press, 1967).

\bibitem{Lumley-Tennekes[1972]}  J.L. Lumley and H.
Tennekes, {\it A First Course in Turbulence}, (MIT Press, 1972).

\bibitem{WReynolds[1987]}  W. C. Reynolds, Fundamentals of turbulence for
turbulence modeling and simulation, in: {\it Lecture Notes for Von Karman
Institute}, AGARD Lecture Note Series, (NATO, New York, 1987) pp.1-66.

\bibitem{Piomelli[1993]}  U. Piomelli, Applications of large eddy simulations
in engineering: an overview, in: {\it Large-eddy Simulation of Complex
Engineering and Geophysical Flows}, ed. B. Galperin and S. A. Orszag
(Cambridge University Press, 1993).

\bibitem{L&M[1996]} M. Lesieur and O. M\'etais, 
Annual Rev. Fluid Mech. {\bf28} (1996) 45.

\bibitem{chen-shan}
S. Y. Chen and X. Shan, Comput. Phys. {\bf 6} (1992) 643.

\bibitem{wang}Lian-Ping Wang, Shiyi Chen, James G. Brasseur and John C. Wyngaard, ``
J. Fluid Mechanics, {\bf 309} (1996) 113.

\bibitem{borue}V. Borue and S. A. Orszag, Europhy. Lett. {\bf 29} (1995) 687.

\bibitem{ccs}N. Cao, S.  Chen and Z. She, 
Phys. Rev. Lett., {\bf 76} (1996) 3711.

\bibitem{cao-chen}N. Cao, S. Chen and G. D. Doolen,
{\em Statistics and Structures of Pressure in Isotropic Turbulence},
Physics of Fluids, in press (1999). 

\bibitem{ashurst}W. T. Ashurst, A. R. Kerstein, R. M. Kerr and 
C. H. Gibson, Phys. Fluids, {\bf 30} (1987) 2343.

\bibitem{k41}
A.N. Kolmogorov, ``The local structure of turbulence in incompressible
viscous fluid for very large Reynolds numbers,'' Dokl. Adad. Nauk SSSR
{\bf 30}, 301 (1941).

\bibitem{k62}
A.N. Kolmogorov, ``A refinement of previous hypotheses concerning the local
structure of turbulence in a viscous incompressible fluid at high Reynolds
number,'' J. Fluid Mech. {\bf 13}, 82 (1962).

\end{thebibliography}
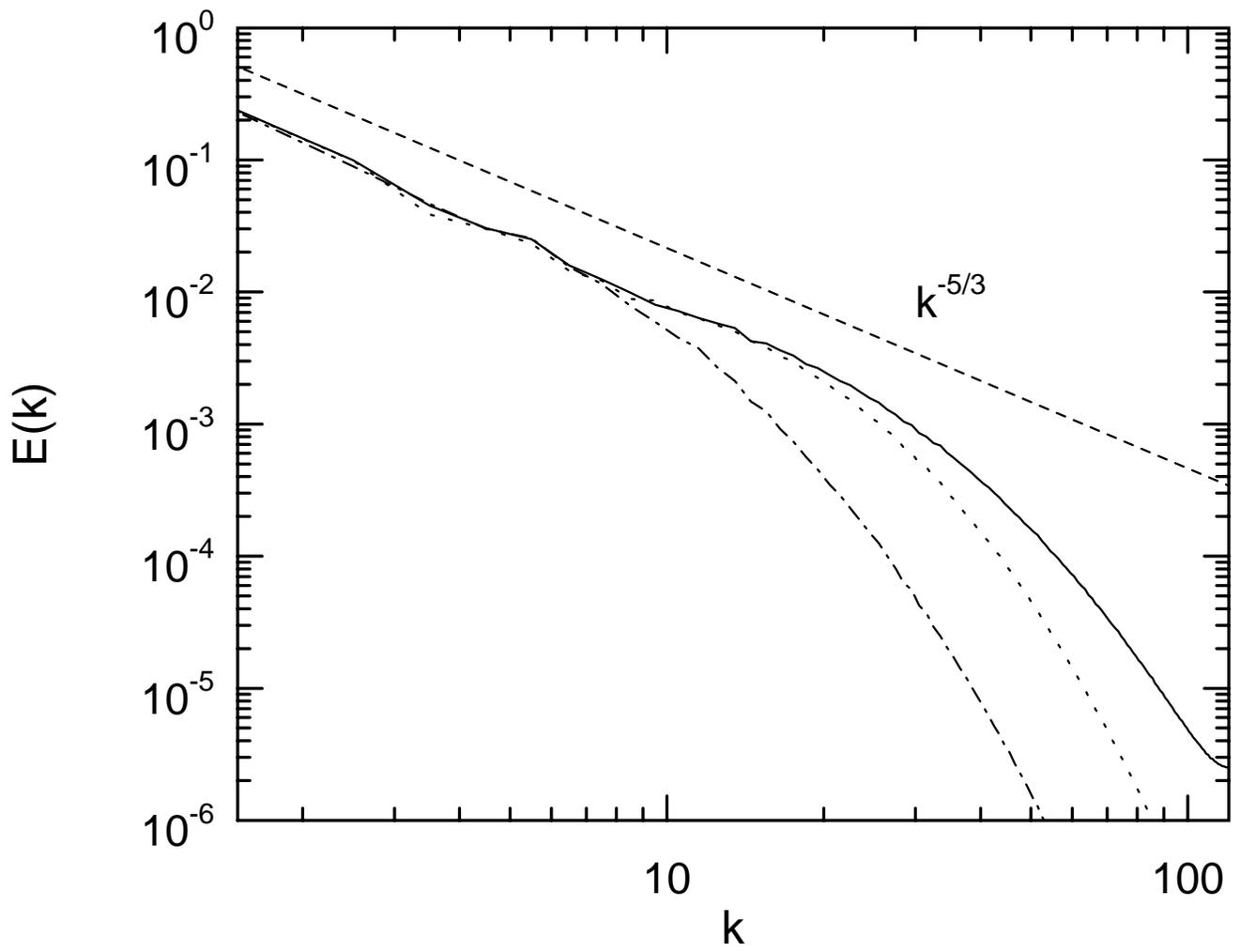
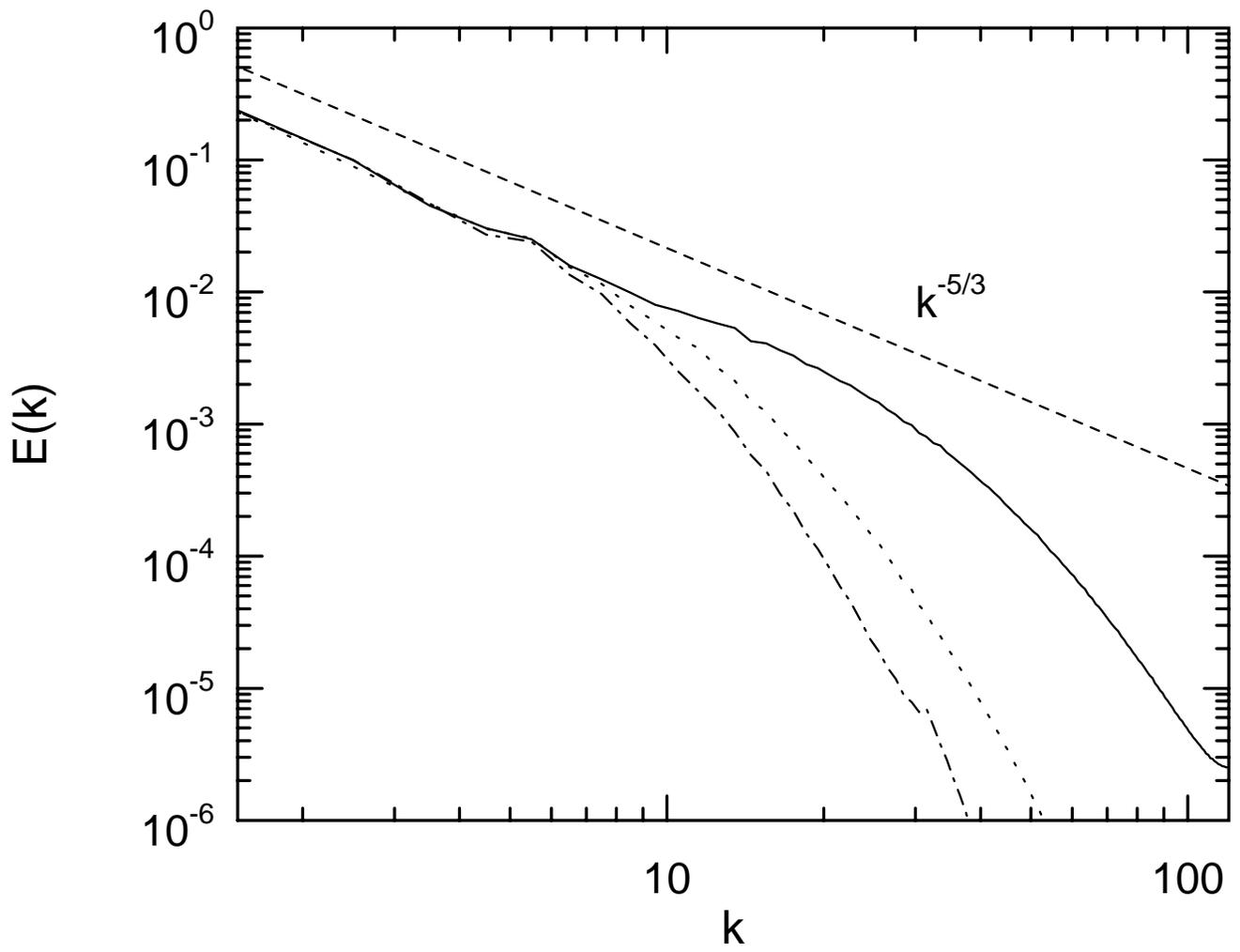
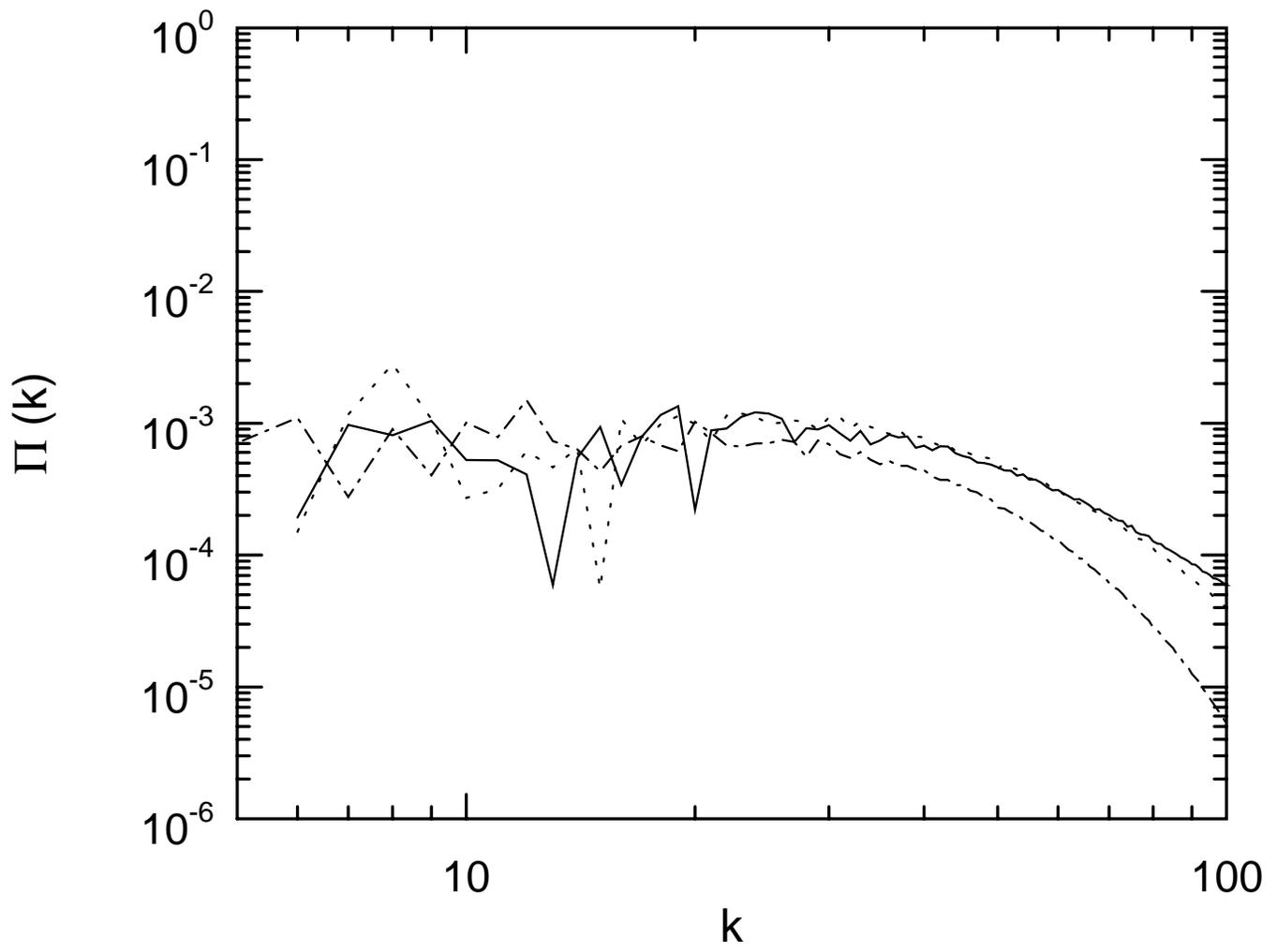
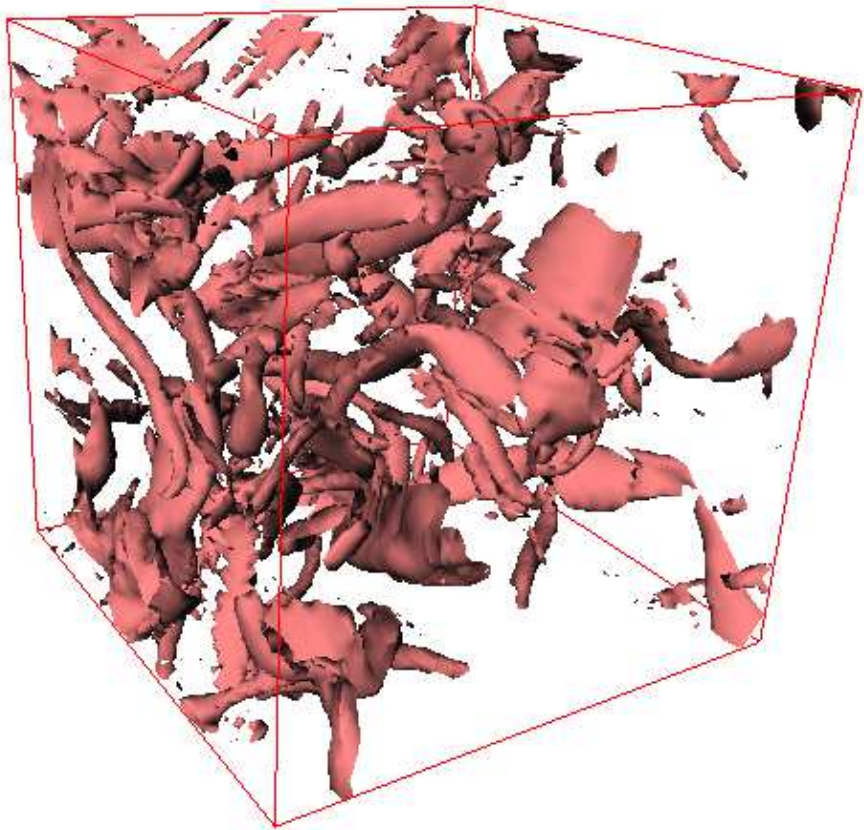
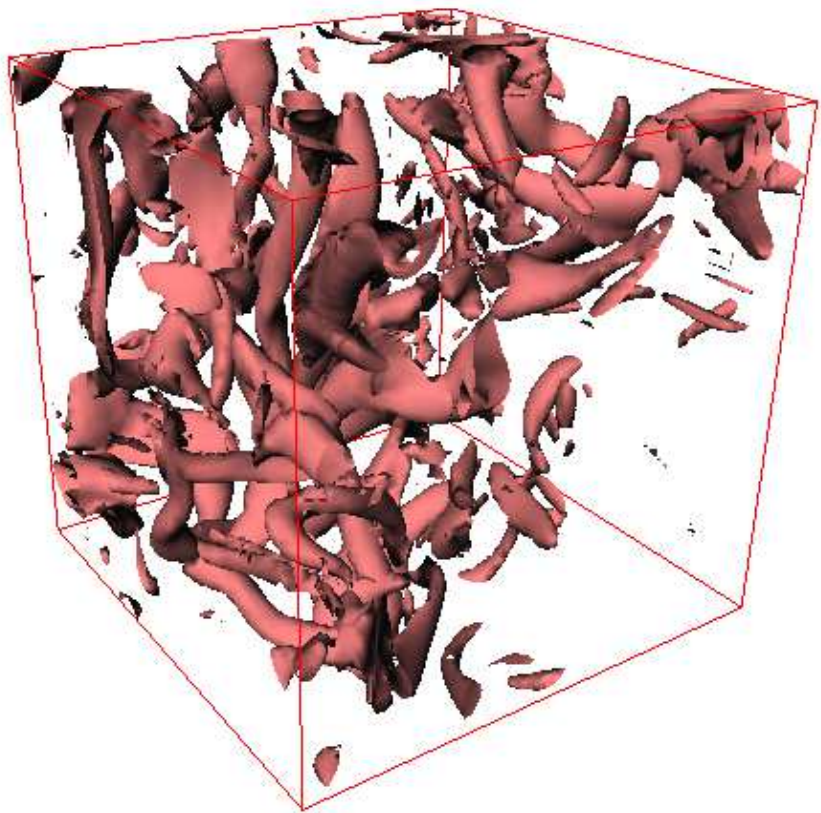
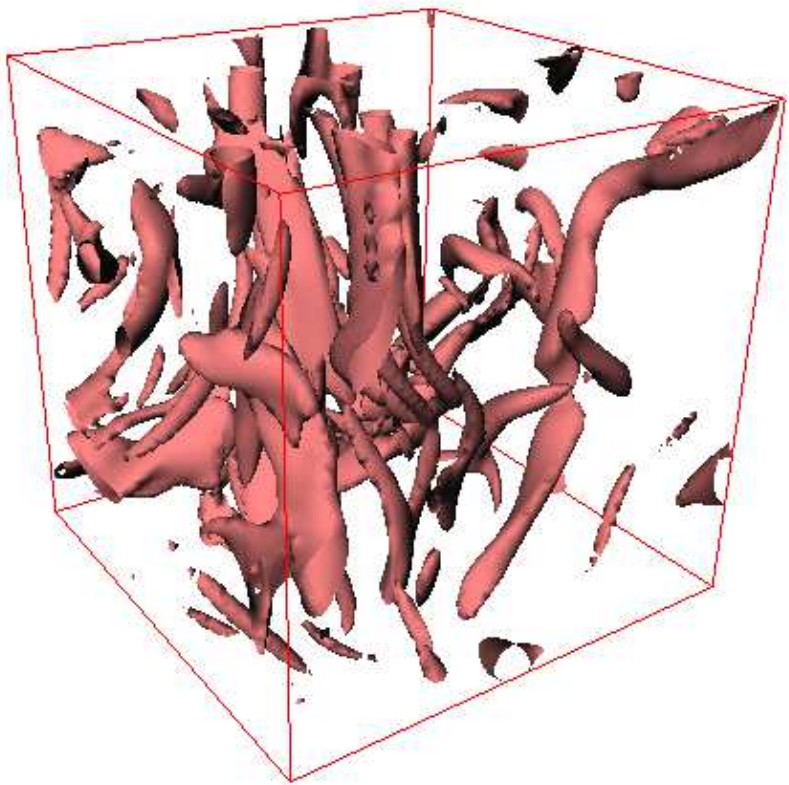
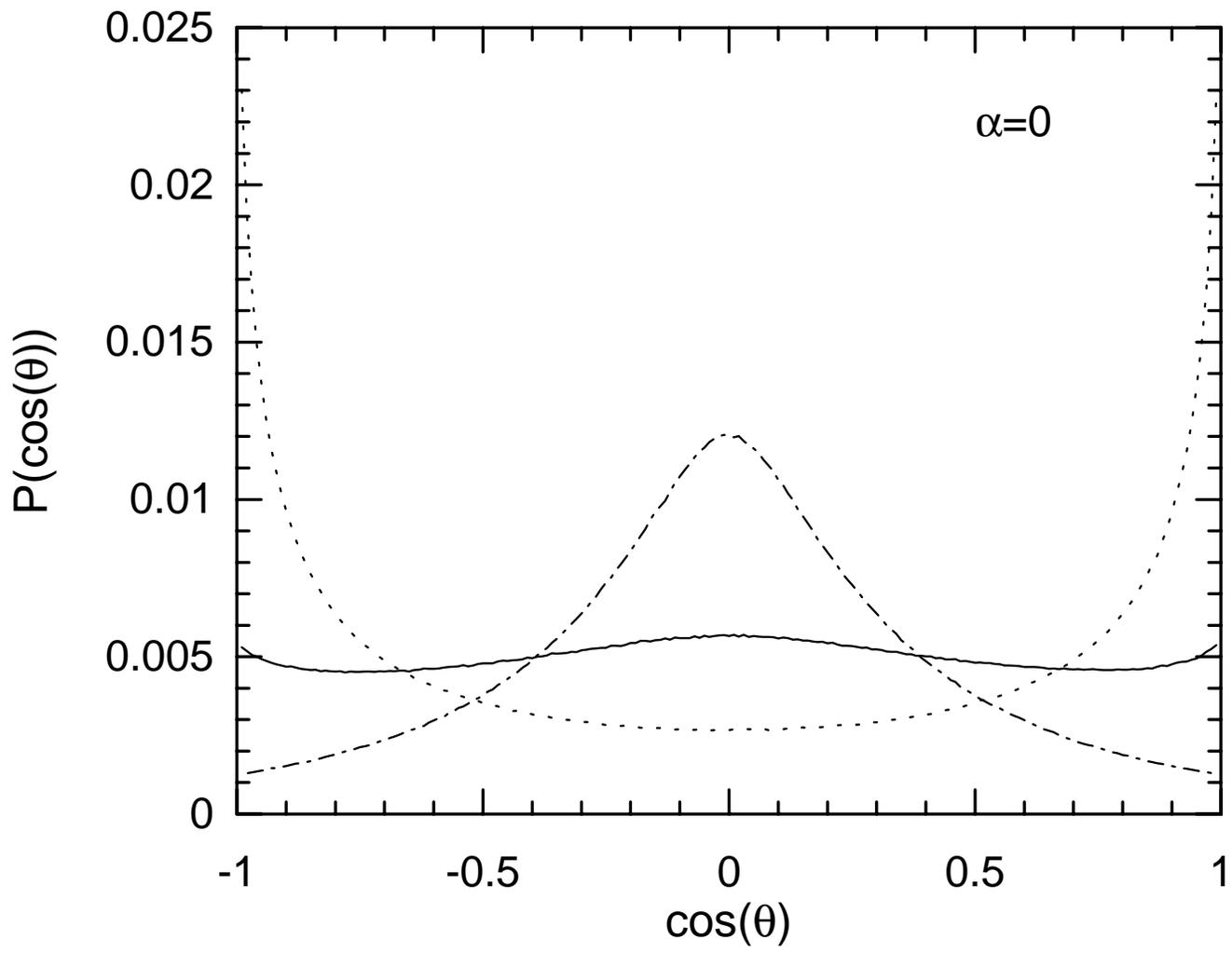
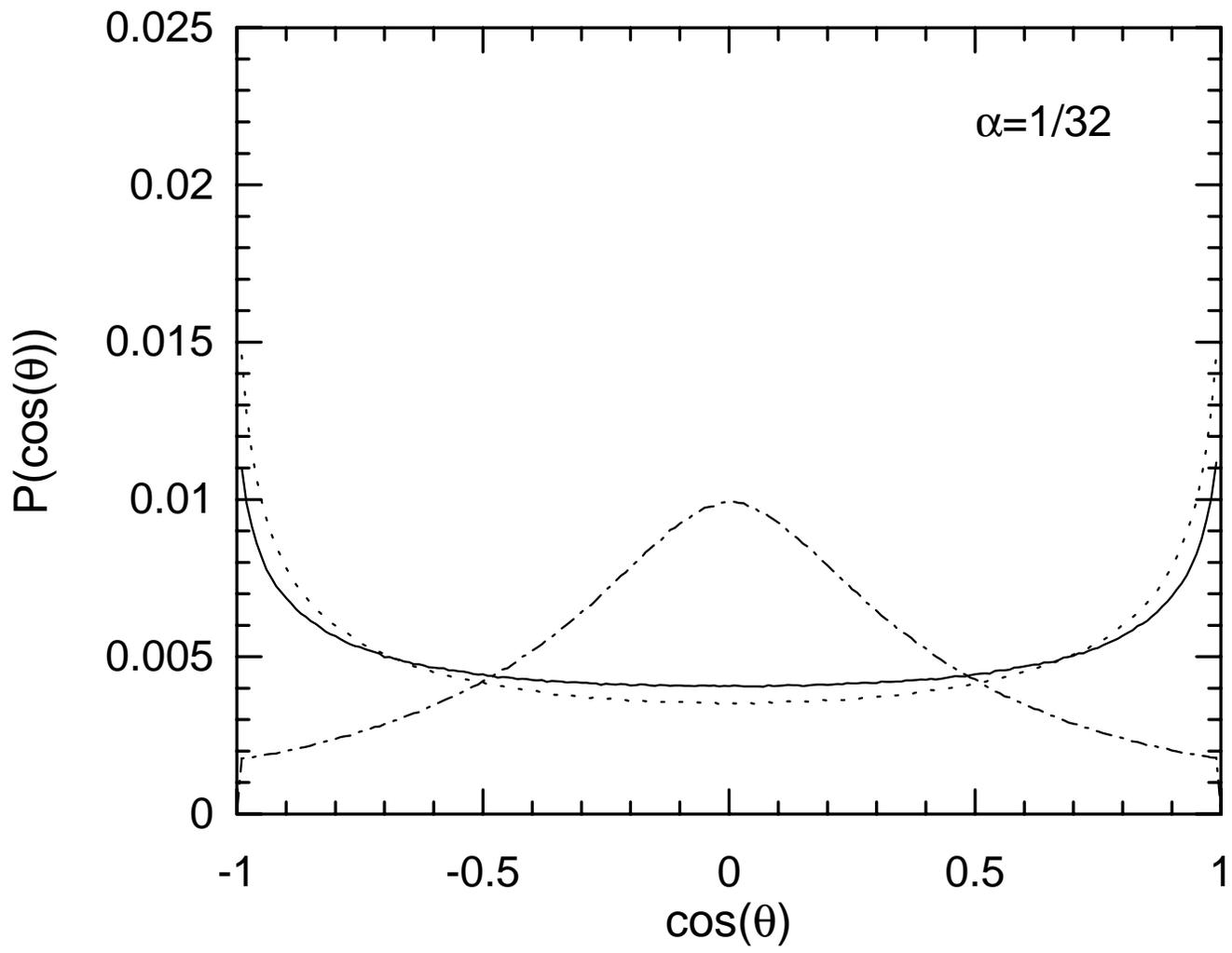
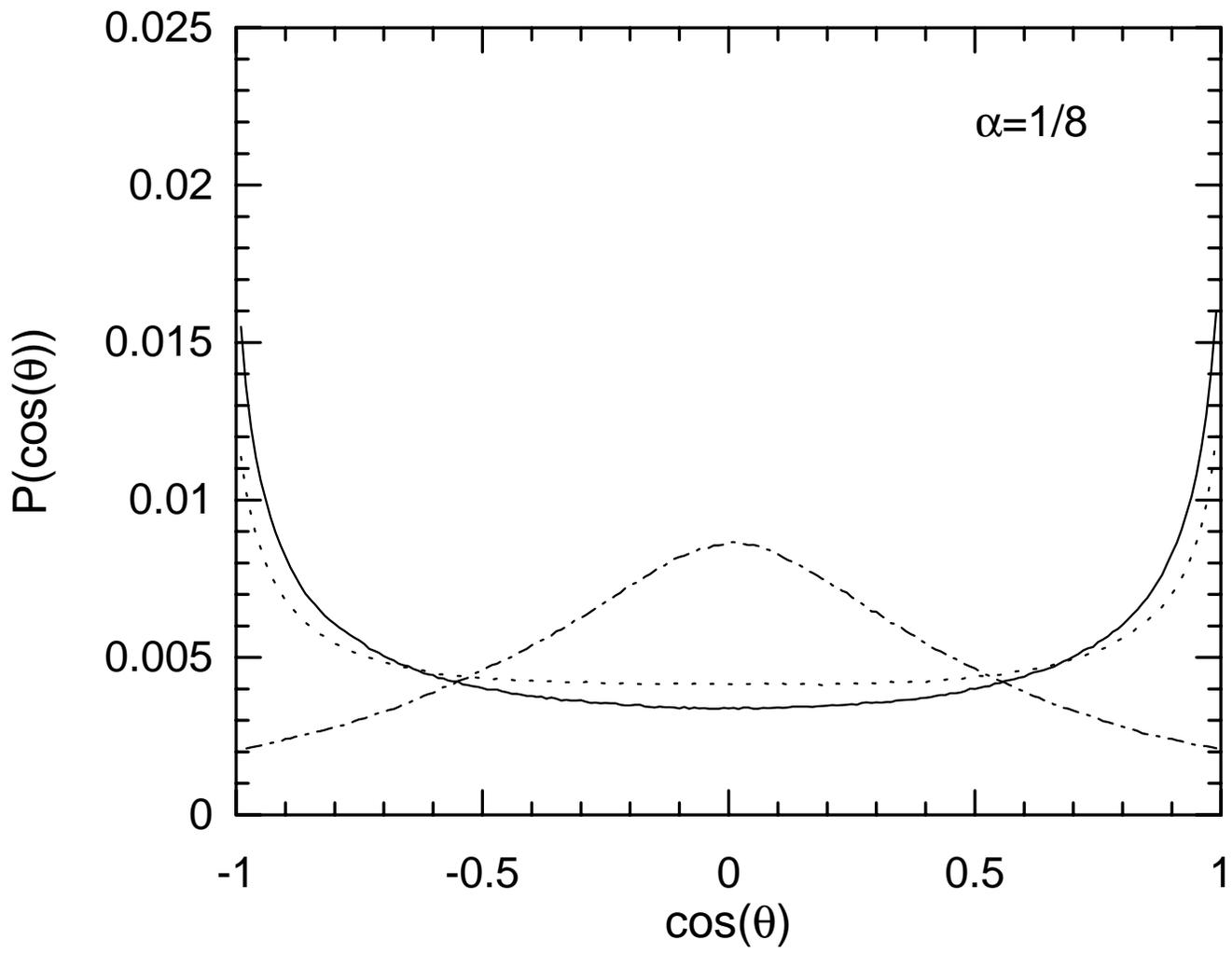
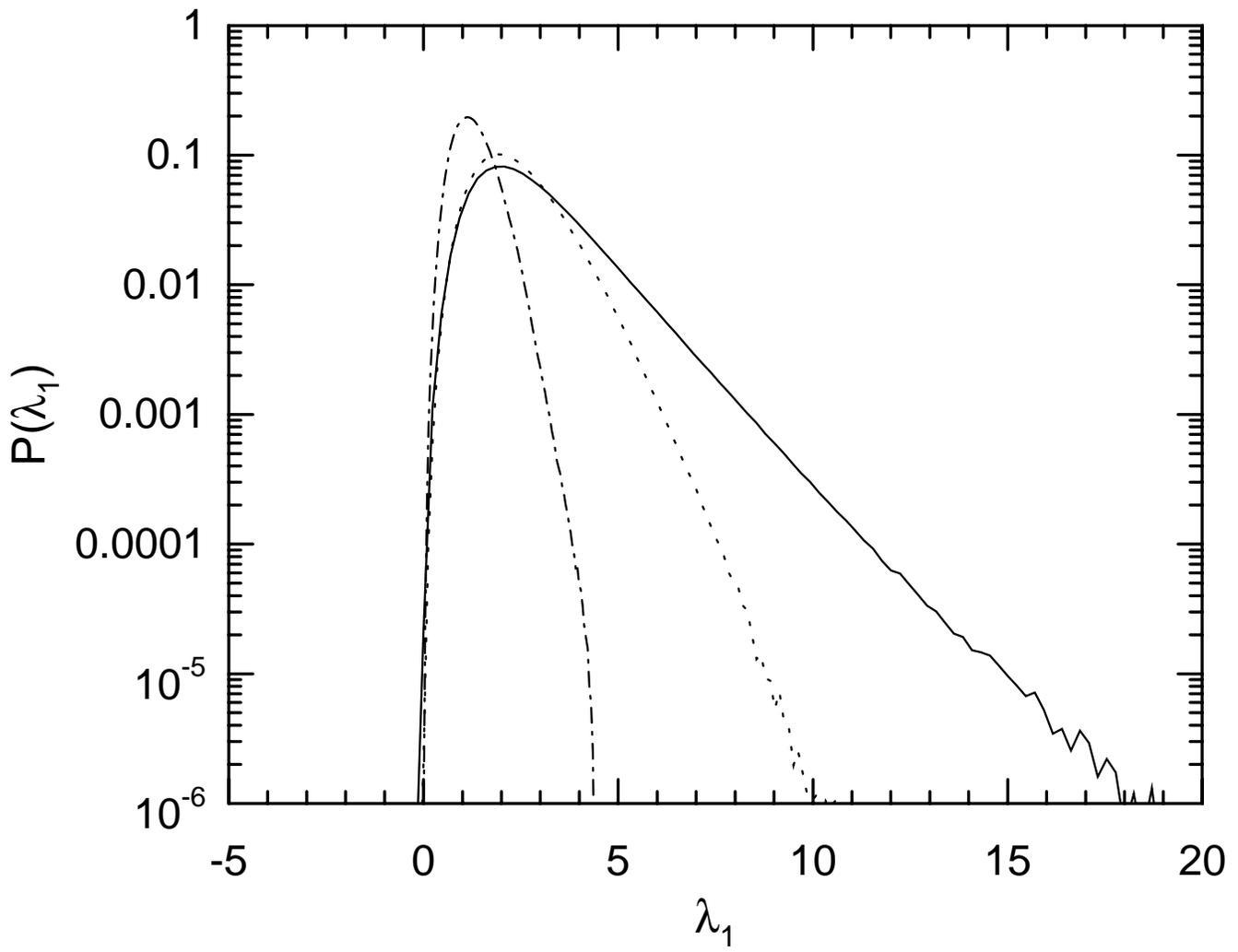
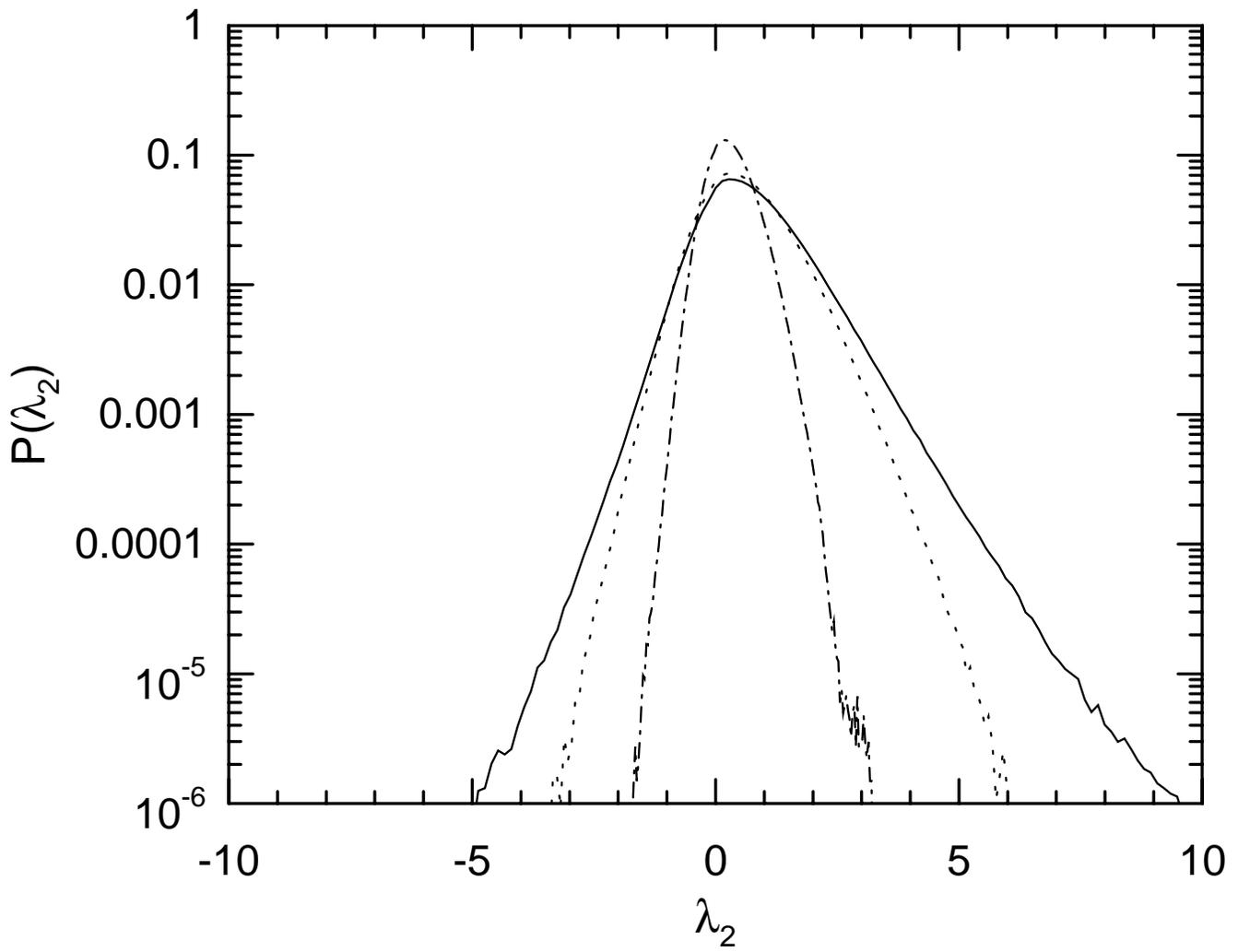
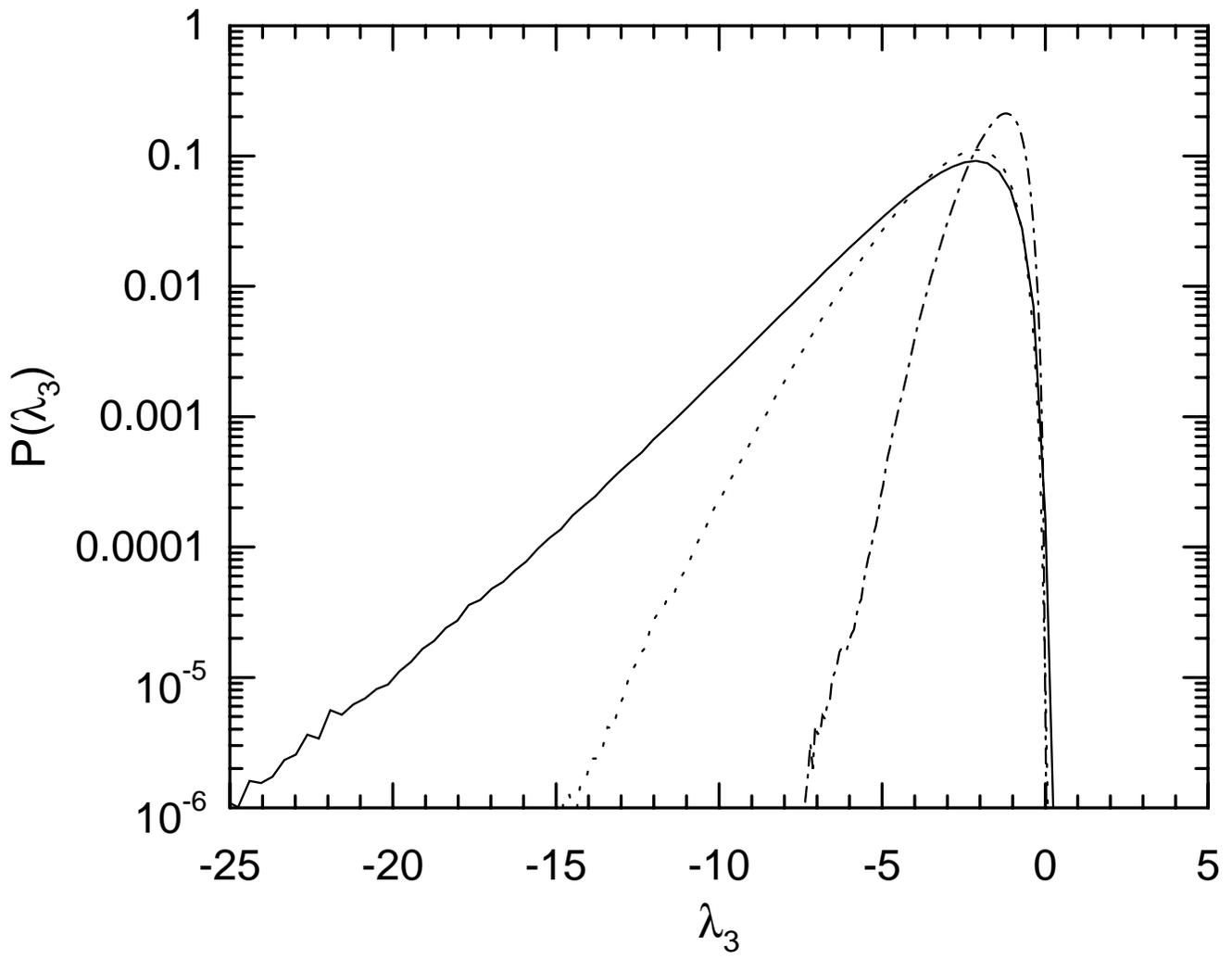
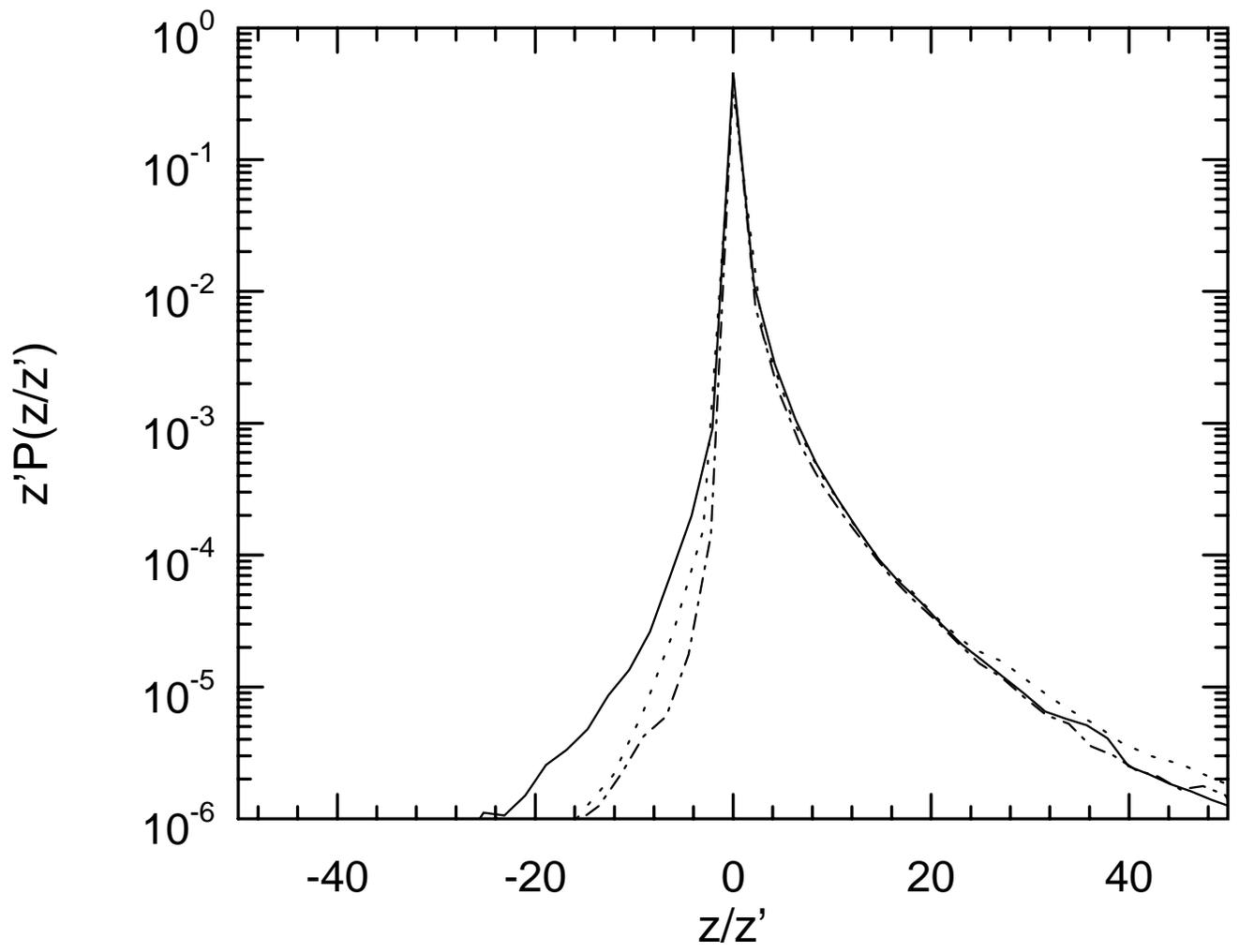
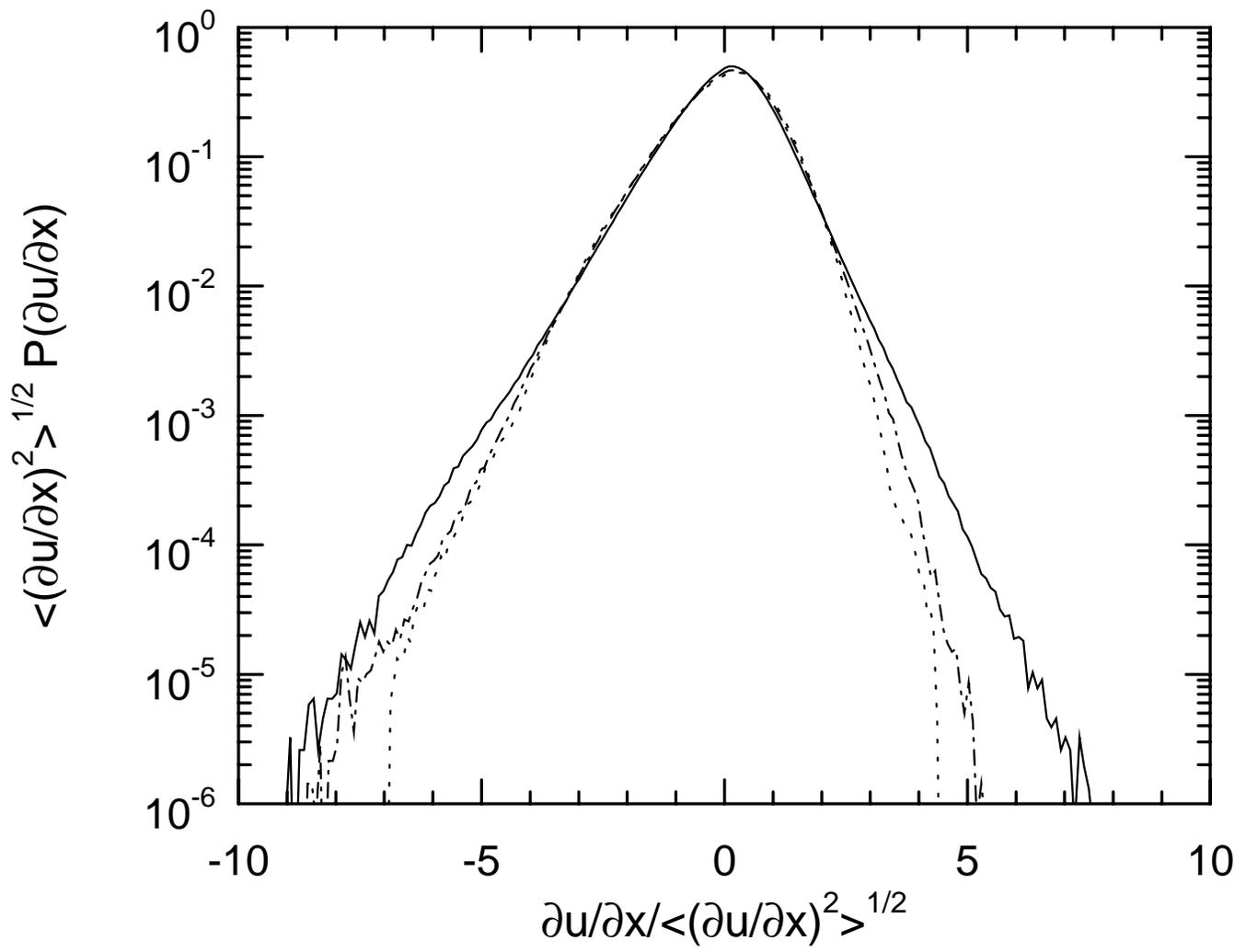
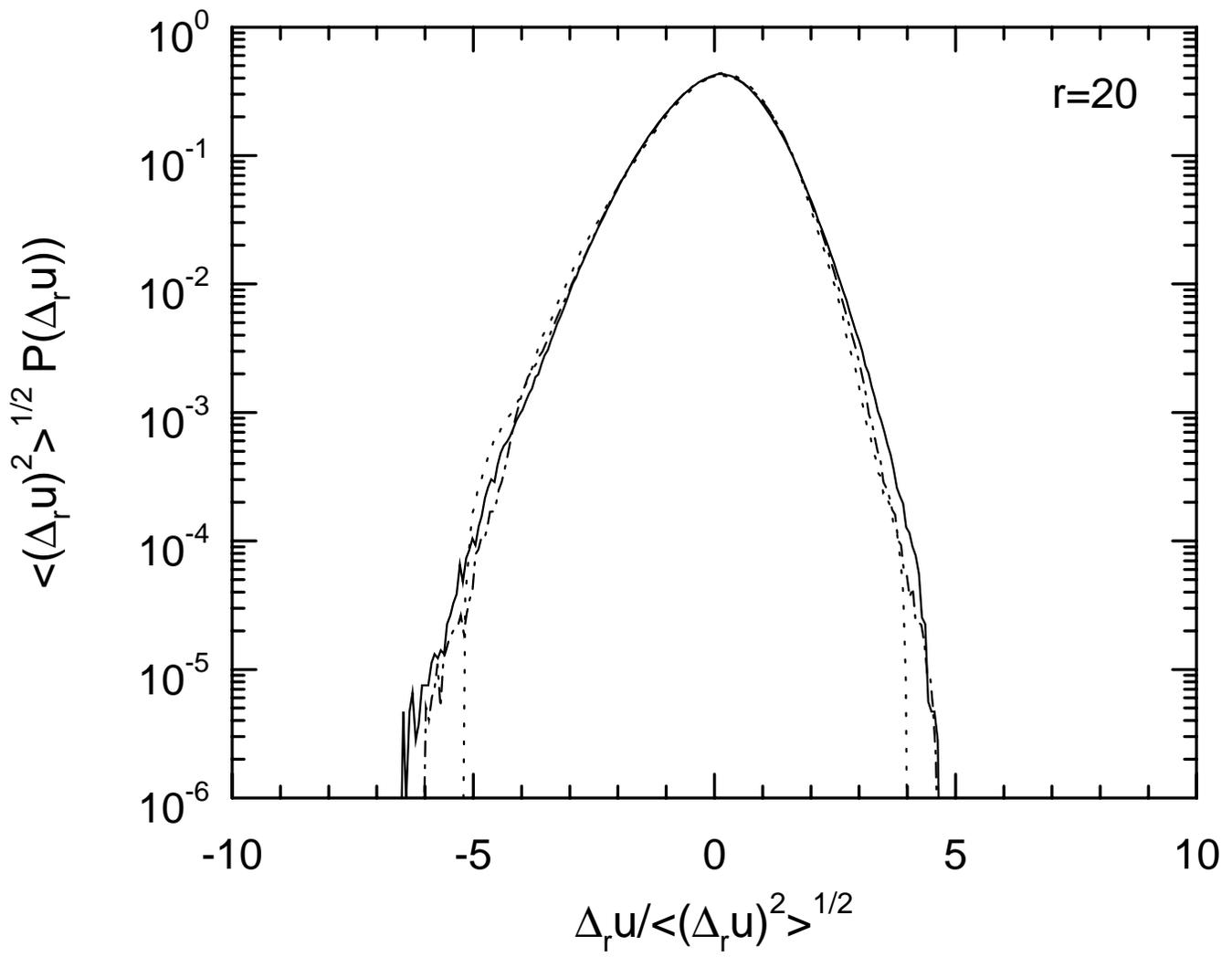
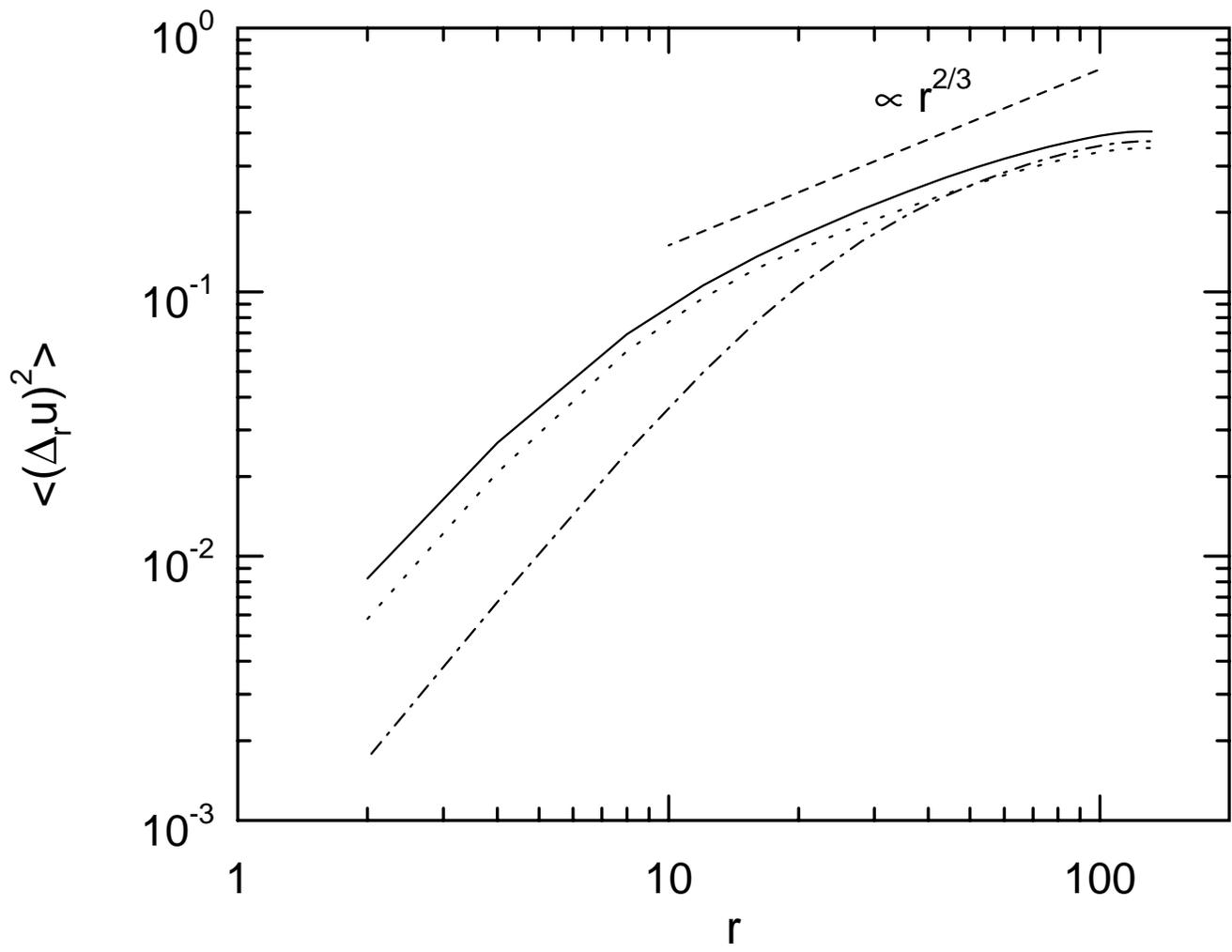
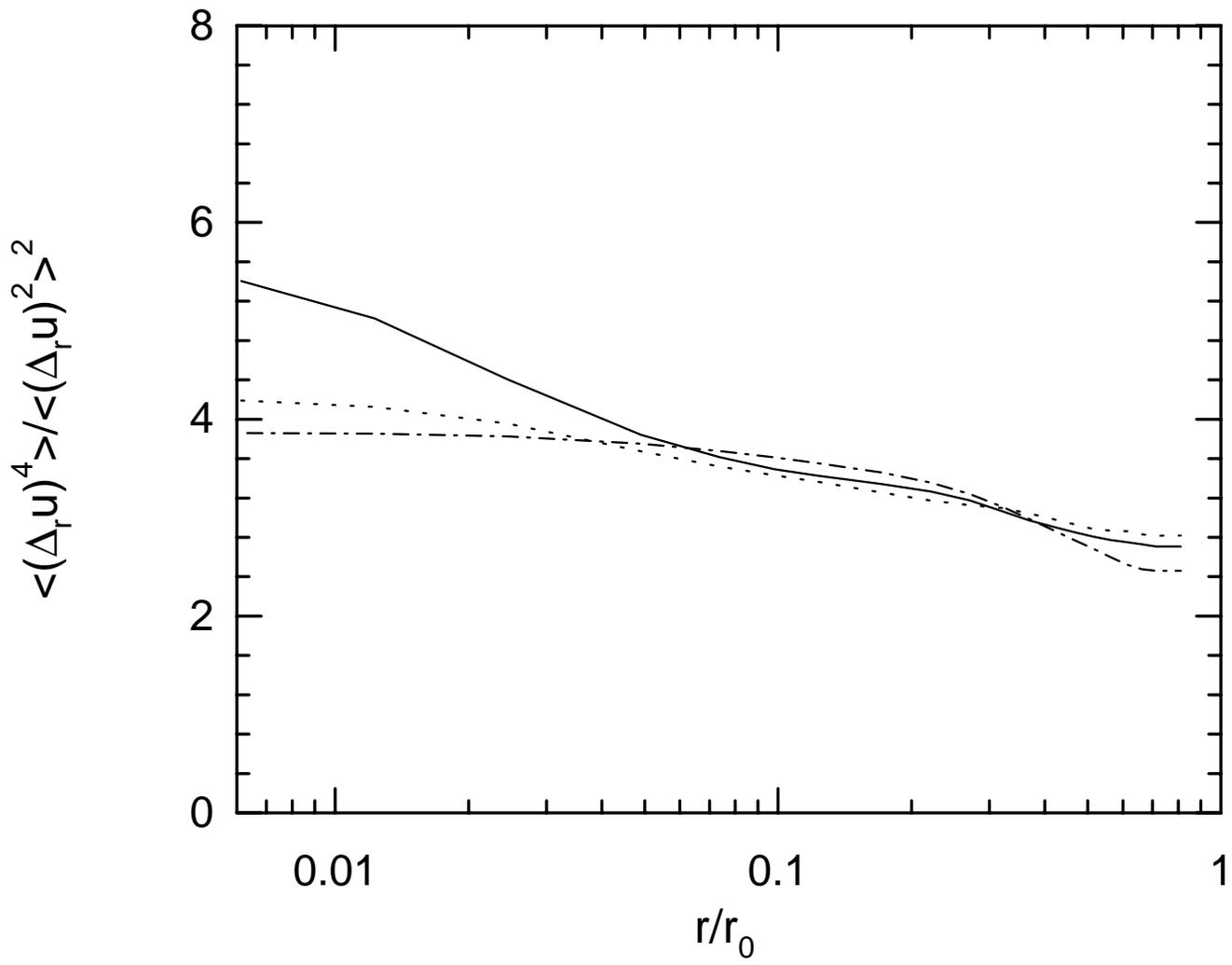
\end{document}